\documentclass[
     11pt,
     a4paper
     ]{article}

\usepackage[
     top=1.5in,
     bottom=1.5in,
     left=1in,
     right=1in
     ]{geometry}
\usepackage{xcolor}
     \definecolor{green_maf}{RGB}{28, 112, 46}
     \definecolor{green_maf_faded}{RGB}{80, 117, 88}
     \definecolor{darkBlue}{RGB}{0,62,133}
     \definecolor{OxfordBlue}{RGB}{0,33,71}
     \definecolor{detail}{RGB}{110,110,110}
     \definecolor{bcite}{RGB}{119, 150, 166}
\usepackage[
     colorlinks=true,
     linkcolor=darkBlue,
     citecolor=darkBlue,
     filecolor=darkBlue,
     urlcolor=darkBlue,
     linktocpage=true
     ]{hyperref}
\usepackage{setspace} 
\usepackage[normalem]{ulem} 
\usepackage{amsmath} 
\usepackage{nccmath} 
\usepackage{amssymb} 
\usepackage{amsthm} 
\usepackage{upgreek} 
\usepackage{simpler-wick} 
\usepackage{datetime} \settimeformat{ampmtime}
\usepackage[new]{old-arrows} 
\usepackage{dsfont} 
\usepackage{bm} 
\usepackage{pbox} 
\usepackage{tikz} 
\usepackage{colortbl} 
\usepackage{mathrsfs} 
\usepackage{mathtools} 
     \mathtoolsset{showonlyrefs} 
\usepackage[utf8]{inputenc} 
\usepackage{fancyhdr} 
\usepackage{changepage} 

\newif\ifcomments
     \commentstrue
\newif\ifdetails
     \detailstrue
\newif\ifmafdetails
     \mafdetailstrue

\newcommand{\no}[1]{{\normalfont\textbf{:}}#1{\normalfont\textbf{:}}} 
\newcommand{\dd}{{\rm d}} 
\newcommand{\boxope}[1]{{\,}\boxed{#1}{\,}} 
\DeclarePairedDelimiter\ket{\lvert}{\rangle} 
\DeclarePairedDelimiterX\braket[2]{\langle}{\rangle}{#1 \delimsize\vert #2} 
\usepackage{multirow}
\usepackage{booktabs}
\usepackage{hhline}
\definecolor{MyRed}{RGB}{255,51,51}
\definecolor{MyBlue}{RGB}{51,153,255}
\definecolor{MyPaleRed}{RGB}{255,204,204}
\definecolor{MyPaleBlue}{RGB}{204,229,255}
\definecolor{verylightgray}{gray}{0.90}
\usepackage[font=small]{caption}

\begin{document}


\title{
$\mathcal{SW}(\tfrac{3}{2},2)$ subsymmetry in
$G_2$, $Spin(7)$
and $\mathcal{N}=2$ CFTs
}

\author{Marc-Antoine Fiset}

\date{\vspace{-5ex}}

\maketitle
\thispagestyle{empty}
\begin{center}
\textit{ \small
Institut f\"ur Theoretische Physik,
ETH Z\"urich,\\
8093 Z\"urich, Switzerland
}\\\vspace{5pt}
\href{mailto:mfiset@phys.ethz.ch}{\texttt{mfiset@phys.ethz.ch}}\\\vspace{10pt}
\end{center}

\abstract{
Spectral flow, spacetime supersymmetry, topological twists, chiral primaries related to marginal deformations, mirror symmetry: these are important consequences of the worldsheet $\mathcal{N}=2$ superconformal symmetry of strings on Calabi--Yau manifolds. To various degrees of certainty, these features were also established when the target is either 7d or 8d with exceptional holonomy $G_2$ or $Spin(7)$ respectively. We show that these are more than mere analogies. We exhibit an underlying symmetry $\mathcal{SW}(\tfrac{3}{2},2)$ making a bridge between the latter cases and K3 target spaces. Reviewing unitary representations of $\mathcal{SW}(\tfrac{3}{2},2)$ leads us to speculate on further roles of this algebra in string theory compactifications and on the existence of topologically twisted versions of $\mathcal{SW}(\tfrac{3}{2},2)$ theories.
}

\newpage

\tableofcontents

\vspace{10pt}

\newpage

\section{Introduction}

The Figueroa-O'Farrill--Schrans (FS) algebra \cite{Figueroa-OFarrill:1990mzn} is the unique family, parametrized by the central charge $c=3\hat{c}/2$, of chiral $\mathcal{N}=1$ superconformal algebras in 2d extended by a primary superfield of conformal weight $2$. We denote it $\mathcal{SW}_{\hat{c}}(\frac{3}{2},2)$ following \cite{Bouwknegt:1992wg}.

This algebra connects interestingly with the geometry of certain real
8d manifolds endowed with a $Spin(7)$-structure,
and certain 7d manifolds with a $G_2$-structure.\footnote{A $\mathcal{G}$-structure on a real $d$-dimensional differentiable manifold, where $\mathcal{G}\subseteq GL(d,\mathbb{R})$ is a Lie group, is defined as a principal subbundle with fibre $\mathcal{G}$ of the frame bundle. The language of $\mathcal{G}$-structures is convenient to discuss the existence of connections on the tangent bundle with holonomy group contained in $\mathcal{G}$. We refer the reader to \cite[ch.\,2]{Joyce2007} for a pedagogical introduction.}
In both cases,
the 2d theory describing string worldsheets mapped in these manifolds is (predicted to flow to) an $\mathcal{N}=1$ superconformal field theory (CFT) enjoying extended chiral symmetries. The vertex algebras formed by these symmetries have been known since Shatashvili and Vafa \cite{Shatashvili:1994zw}. We refer to them as $\text{SV}^{Spin(7)}$ and $\text{SV}^{G_2}$.\footnote{For type II, there is a copy of these algebras for each chirality. For heterotic strings, only one acts on the supersymmetric side \cite{Melnikov:2017yvz}, while the ordinary Virasoro algebra acts on the non-supersymmetric side.} 
Their rescaled central charge $\hat{c}$ matches the dimension of the manifold, respectively $\hat{c}=8$ and $\hat{c}=7$.

The pair of generators $(X, M)$ used in \cite{Shatashvili:1994zw} is slightly unnatural in that it does not form a primary superfield (in either algebra). In the $Spin(7)$ algebra, it was however noticed \cite{Figueroa-OFarrill:1996tnk} that $(X, M)$ and the superconformal multiplet $(\tfrac{1}{2}G,T)$ can be combined into a weight $2$ superprimary. By uniqueness of the FS algebra, 
the $Spin(7)$ and FS algebras are thus identical at central charge $12$, or $\hat{c}=8$:
\begin{equation} \label{eq:FS_Spin7}
\text{SV}^{Spin(7)} = \mathcal{SW}_{8}(\tfrac{3}{2},2)\qquad(\hat{c}=8) \,.
\end{equation}

Meanwhile in the $G_2$ Shatashvili--Vafa algebra,
a proper subalgebra is generated by the fields $T,G,X,$ and $M$ \cite{Noyvert:2002mc}. 
It turns out again that a weight $2$ superprimary can be identified by combining these generators and, by uniqueness, the subalgebra is again identified with FS, now at a different central charge:
\begin{equation} \label{eq:FS_G2}
\text{SV}^{G_2} \supset \mathcal{SW}_{7}(\tfrac{3}{2},2) \qquad (\hat{c}=7)\,.
\end{equation}

It is a remarkable fact---and in our opinion insufficiently appreciated---that the \textit{same} algebra FS arises from string dynamics on \textit{both} $Spin(7)$ and $G_2$-structure manifolds, which otherwise appear as exceptional and isolated mathematical objects. It is as if a hidden uniformity existed for these exceptional geometries once stringy corrections are taken into account. 
This note takes a closer look at this common denominator in the physics of these models.

After rediscovering \cite{Figueroa-OFarrill:1990mzn, Blumenhagen:1991nm, Blumenhagen:1992vr} the $Spin(7)$ and $G_2$ algebras, Shatashvili and Vafa described, in varying depth, their many surprising similarities with $\mathcal{N}=(2,2)$ CFTs:
\begin{itemize}
\item analogue of chiral or anti-chiral primaries,
\item discrete notion of spectral flow,
\item relations with marginal deformations,
\item exotic topological twist,
\item spacetime supersymmetry.
\end{itemize}
We shall stress in this note (sections \ref{sec:Properties} and \ref{sec:Twists}) that these are actually features of the FS subsymmetry $\mathcal{SW}_{\hat{c}}(\tfrac{3}{2},2)$\,.

We shall show furthermore that
there is nothing special about $\hat{c}=8$ and $7$
as far as the interesting properties are concerned, leading us to consider the whole FS family at once. We review in section~\ref{sec:FSunitarity} the constraints from unitarity \cite{Gepner:2001px} and present the discrete sequence of central charges admitting unitary representations. In addition to $\hat{c}=8$ and $7$, this sequence includes the integer values $6$ and $5$, and it accumulates at $\hat{c}=4$. The critical superstring dimension $10$ also appears as the maximal permitted value of $\hat{c}$. This suggests that the FS symmetry, underpinning strings in 8d $Spin(7)$ and 7d $G_2$ manifolds, carries over to other types of manifolds of integer dimension as well.

We confirm this intuition in the case $\hat{c}=4$ in section~\ref{sec:N=2}. We show that the FS algebra with $\hat{c}=4$ arises as a subalgebra of the $\mathcal{N}=2$ superconformal Virasoro algebra
at the same central charge:
\begin{equation} \label{eq:FS_Vir}
\text{\normalfont Vir}^{\mathcal{N}=2} \supset \mathcal{SW}_{4}(\tfrac{3}{2},2) \qquad (\hat{c}=4) \,.
\end{equation}

This contact with the extensively studied $\mathcal{N}=2$ case promises to clarify some of the more opaque aspects of FS theories---for example how to twist these $\mathcal{N}=1$ theories with trivial R-symmetry---and their applications. We make steps in this direction in section~\ref{sec:Twists} by generalizing to all unitary FS theories the arguments in \cite{Shatashvili:1994zw} in favour of a topological twist. We argue that this twist is analogous to the $(+)$ twist of $\mathcal{N}=2$, and we suggest that an analogue of the $(-)$ twist may also be worth investigating in FS theories. We finally comment on how one might be able to make rigorous these twists by invoking conformal block decompositions, generalizing ideas in \cite{deBoer:2005pt}.

The geometric relevance of the FS algebra for $\hat{c}=5$, $6$ and $10$, if any, remains intriguing and open. In the case $\hat{c}=6$, we rule out in section~\ref{sec:N=2} what might have been the natural expectation: the $\mathcal{N}=2$ superconformal algebra at $\hat{c}=6$ \emph{does not} contain any FS sublagebra, despite their many similarities. FS also does not fit inside the so-called Odake algebra \cite{Odake:1988bh} associated to Calabi--Yau $3$-folds. There curiously seems to exist two genuinely distinct algebras at $\hat{c}=6$ with very similar properties. Their comparison deserves further investigation.

\section{$\mathcal{SW}_{\hat{c}}(\frac{3}{2},2)$, unitarity and hidden sectors} \label{sec:FSunitarity}

We start by introducing the FS algebra $\mathcal{SW}_{\hat{c}}(\frac{3}{2},2)$ \cite{Figueroa-OFarrill:1990mzn, Figueroa-OFarrill:1990tqt}.
Let $(\tfrac{1}{2}G, T)$ generate a copy of the $\mathcal{N}=1$ superconformal algebra with central charge $c=3\hat{c}/2$, given in operator form by the OPEs
\begin{align}
\wick{\c T(z) \c T(w)} &= \frac{c/2}{(z-w)^4}+\frac{2 T(w)}{(z-w)^2}+\frac{T'(w)}{z-w} \,, 
\label{eq:TT}\\
\wick{\c T(z) \c G(w)} &= \frac{3/2}{(z-w)^2} G(w)+\frac{G'(w)}{z-w} \,, 
\label{eq:TG}
\\
\wick{\c G(z) \c G(w)} &= \frac{2c/3}{(z-w)^3}+\frac{2 T(w)}{z-w} \,.
\label{eq:GG}
\end{align}
The contraction bracket symbolizes that we retain only singular terms and prime ($'$) is the derivative with respect to a holomorphic coordinate $z\in\mathbb{C}$.

Let $(W,U)$ form an $\mathcal{N}=1$ superprimary where the leading component is bosonic with conformal weight $2$ and $U$ is fermionic with weight $5/2$:
\begin{align}
\wick{\c T(z) \c W(w)} &= \frac{2 W(w)}{(z-w)^2}+\frac{W'(w)}{z-w} \,, \label{eq:TW}\\
\wick{\c T(z) \c U(w)} &= \frac{5/2}{(z-w)^2}U(w)+\frac{U'(w)}{z-w} \,, \label{eq:TU}\\
\wick{\c G(z) \c W(w)} &= \frac{U(w)}{z-w} \,,
\label{eq:GW}\\
\wick{\c G(z) \c U(w)} &= \frac{4 W(w)}{(z-w)^2}+\frac{W'(w)}{z-w} \,.
\label{eq:GU}
\end{align}
The remaining OPEs are fixed by associativity bootstrap. The freedom left after enforcing the Jacobi-like identities \cite{Thielemans:1994er} is parametrized by the central charge and the normalization of the superprimary. In the standard normalization \cite{Figueroa-OFarrill:1990mzn,Gepner:2001px} where
\begin{equation}
\wick{\c W(z) \c W(w)} = \frac{c/2}{(z-w)^4}+\text{less singular}\,,
\end{equation}
the case $c=15$ spuriously appears to be singular and square roots must be artificially introduced. We choose here a non-standard normalization to present the remaining OPEs more transparently as follows. (Arguments $w$ are sometimes omitted.)
\begin{align}
\wick{\c W(z) \c W(w)} &=
\frac{ c \mu/2}{(z-w)^4}
+\frac{2(\mu T-\nu W)}{(z-w)^2}
+\frac{\mu T'-\nu W'}{z-w}
\label{eq:WW}\\
\wick{\c W(z) \c U(w)} &=
-\frac{3 \mu G}{(z-w)^3}
-\frac{\mu G'+ \nu U}{(z-w)^2}
+\frac{\boxope{WU}_1}{z-w}
\label{eq:WU}\\
\wick{\c U(z) \c U(w)} &=
-\frac{2 c \mu}{(z-w)^5}
+\frac{2(2 \nu W-5\mu T)}{(z-w)^3}
+\frac{2 \nu W'-5 \mu T'}{(z-w)^2} 
+\frac{\boxope{UU}_1}{z-w}
\label{eq:UU}
\end{align}
We have used the shorthands
\begin{equation}
\mu{\;}{{{=}}}{\;}(15-c)(4c+21)\,,\qquad \nu{\;}{{{=}}}{\;}5c+6 \,,
\end{equation}
and defined the fields
\begin{align}
\boxope{WU}_1 &{\;}{{{=}}}{\;} 54 (c-15) \no{TG}+54 \no{GW}+(c-15)^2 G''-2 (c+12) U' \,,\\
\boxope{UU}_1 &{\;}{{{=}}}{\;} 3 \Big[        (c-15)\big(  9 \no{G'G}+36 \no{TT}+(2c-3)T''  \big)      -18 \no{GU}+36 \no{TW}+(c-6) W''\Big] \,. \nonumber
\end{align}
Colons ($\no{~}$) represent normal ordering.

We immediately remark that $T$ and $W$ close a subalgebra of type $\mathcal{W}(2,2)$, namely Virasoro extended by a weight $2$ primary.
In fact, eq.\,\eqref{eq:WW} is precisely the $WW$ OPE in a general $\mathcal{W}(2,2)$ algebra, with the exception that the constants $\mu$ and $\nu$, above related to the central charge, are in general left undetermined (with $\nu^2$ to $\mu$ being the physically meaningful ratio). It was recognized already by Zamolodchikov \cite{Zamolodchikov:1986gt} that $\mathcal{W}(2,2)$ is almost always equivalent to the tensor product 
of two commuting Virasoro algebras,
\begin{equation}
\wick{\c T^< (z) ~ \c T^> (w)} = 0 \,.
\end{equation}
The only exception is when $\nu^2+4\mu=0$, corresponding to $c=-12$ within the FS algebra. The explicit linear combinations bearing this fact are \cite{Gepner:2001px}
\begin{equation} \label{eq:T<>}
T^<{\;}{{{=}}}{\;}\frac{c^<}{c}
T-\frac{1}{3|c+12|}W \,,
\qquad\qquad
T^>{\;}{{{=}}}{\;}\frac{c^>}{c}
T+\frac{1}{3|c+12|}W \,,
\end{equation}
and the central charges are
\begin{equation} \label{eq:c<>}
c^<{\;}{{{=}}}{\;}\frac{c}{2}\left(1 - \frac{5c+6}{3|c+12|} \right)\,,
\qquad\qquad
c^>{\;}{{{=}}}{\;}\frac{c}{2}\left(1 + \frac{5c+6}{3|c+12|} \right)\,.
\end{equation}
We shall refer to $T^<$ and $T^>$ as the hidden \textit{small} and \textit{large} Virasoro operators respectively. We have $c=c^< +c^>$  and \eqref{eq:T<>} inverts into
\begin{equation}
T=T^<+T^>\,,\qquad
W=\frac{3}{2}|c+12|(T^>-T^<)-\frac{5c+6}{2}T \,.
\end{equation}

\begin{figure}[htbp]
\begin{center}
\begin{tikzpicture}{scale=1}
\draw[->] (-1,0) -- (10.5,0) node[right] {$\hat{c}$};
\draw[->] (0,-1) -- (0,5.5) node[above] {$c^<$};

\draw[scale=1,domain=-0.25:10.5,smooth,variable=\d,black] plot ({\d},{5*\d*(10-\d)/2/(\d+8)});

\node at (0.7,5.15) {\footnotesize{$q\rightarrow\infty$}};
\node at (0.7,4.6077) {\footnotesize{$q=12$}};
\node at (0.7,4.0857) {\footnotesize{$q=6$}};
\node at (0.7,3.3) {\footnotesize{$q=4$}};
\node[fill=white] at (0.7,2.3) {\footnotesize{$q=3$}};
\node at (0.7,-0.2) {\footnotesize{$q=2$}};

\draw[scale=1,domain=-1.5:4,dashed,variable=\y,gray]  plot ({\y},{5});
\draw[scale=1,domain=-1.5:5,dashed,variable=\y,gray]  plot ({\y},{5*(1-6/12/(12+1))});
\draw[scale=1,domain=-1.5:6,dashed,variable=\y,gray]  plot ({\y},{5*(1-6/6/(6+1))});
\draw[scale=1,domain=-1.5:7,dashed,variable=\y,gray]  plot ({\y},{5*(1-6/4/(4+1))});      
\draw[scale=1,domain=-1.5:8,dashed,variable=\y,gray]  plot ({\y},{5*(1-6/3/(3+1))});

\node at (-0.7,5.15) {\footnotesize{$1$}};
\node at (-0.7,4.6077) {\footnotesize{$25/26$}};
\node at (-0.7,4.0857) {\footnotesize{$6/7$}};
\node at (-0.7,3.3) {\footnotesize{$7/10$}};
\node at (-0.7,2.3) {\footnotesize{$1/2$}};
\node at (-0.7,-0.2) {\footnotesize{$0$}};

\node at (7.5,4.3) {\Large $\nwarrow$};

\draw[scale=1,domain=-0.2:5,dashed,variable=\y,gray]  plot ({4},{\y});
\draw[scale=1,domain=-0.2:4.8077,dashed,variable=\y,gray]  plot ({5},{\y});
\draw[scale=1,domain=-0.2:4.2857,dashed,variable=\y,gray]  plot ({6},{\y});
\draw[scale=1,domain=-0.2:3.5,dashed,variable=\y,gray]  plot ({7},{\y});
\draw[scale=1,domain=-0.2:2.5,dashed,variable=\y,gray]  plot ({8},{\y});
\draw[scale=1,dashed,gray] (10,-0.2)--(10,0.4);

\node at (4,5.0) {$\bullet$};
\node at (5,4.8077) {$\bullet$};
\node at (6,4.2857) {$\bullet$};
\node at (7,3.5) {$\bullet$};
\node at (8,2.5) {$\bullet$};
\node at (10,0) {$\bullet$};

\node at (4,-0.4) {\footnotesize{4}};
\node at (5,-0.4) {\footnotesize{5}};
\node at (6,-0.4) {\footnotesize{6}};
\node at (7,-0.4) {\footnotesize{7}};
\node at (8,-0.4) {\footnotesize{8}};
\node at (10,-0.4) {\footnotesize{10}};
\end{tikzpicture}
\caption{Hidden small central charge $c^<$ vs $\hat{c}$. Integer values of $\hat{c}$ in the $\nwarrow$ sequence compatible with unitarity are identified with $\bullet$.}
\label{fig:c^</c_hat}
\end{center}
\end{figure}
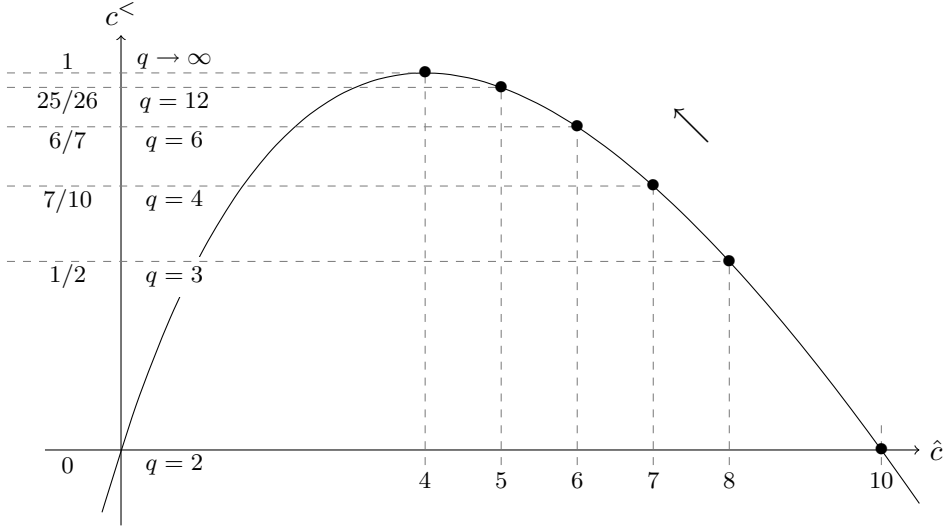

Unitary representations of an algebra must also provide unitary representations of all its subalgebras \cite{Gepner:2001px}. The hidden small Virasoro subalgebra turns out to highly constrain the FS representation theory. Note indeed that $c^<$, regarded as a function of $c > -12$ (see fig.\,\ref{fig:c^</c_hat}), is negative everywhere except between its zeros, which are at $c=0$ and $c=15$. Between these zeros, we have $0<c^<\leq 1$. By the non-unitarity proof of \cite{Friedan:1983xq, Friedan:1986kd}, we conclude \cite{Gepner:2001px} that the FS chiral algebra does not admit unitary representations away from the special values of $c$ such that $T^<$ defines a Virasoro unitary minimal model.
\begin{equation}
\text{Unitarity}~\Rightarrow \qquad c^< = 1-\frac{6}{q(q+1)}\qquad \text{for some integer }q=2,3,\ldots,\infty\,.
\end{equation}

Solving for $c$ or $\hat{c}$ in terms of $q$ yields two sequences accumulating as $q\rightarrow\infty$ at the point $(\hat{c},c^<)=(4,1)$. One ascends from the left ($\nearrow$) and one from the right ($\nwarrow$). The accumulation point $c^<=1$ allows unitary representations \cite{Gepner:2001px} and so should be included.
\begin{align}
&\nearrow\qquad \hat{c}=\frac{4(q-2)}{q+1} \qquad c^>=\frac{(q-2)(5q-3)}{q(q+1)} \qquad q=2, 3, \ldots, \infty \\
&\nwarrow\qquad \hat{c}=\frac{4(q+3)}{q} \qquad c^>=\frac{(q+3)(5q+8)}{q(q+1)} \qquad q=2, 3, \ldots, \infty \label{eq:nwsequence}
\end{align}

The ``large'' Virasoro subalgebra places no extra unitarity constraints.\footnote{Over $c\geq 0$, the large central charge $c^>$ monotonously increases as a function of $\hat{c}$ from $(\hat{c},c^>)=(0,0)$ and it crosses the point $(\hat{c},c^>)=(1,1)$ before increasing further. These two points correspond exactly to the consecutive values $q=2$ and $q=3$ in the $\nearrow$ sequence already found, so the range $0< \hat{c} < 1$ is fully excluded by unitarity.}

Let us focus on the $\nwarrow$ sequence. For $q=3$ ($\hat{c}=8$), the FS algebra is exactly the $Spin(7)$ Shatashvili--Vafa algebra \cite{Figueroa-OFarrill:1996tnk}; see \eqref{eq:FS_Spin7} in the introduction. The superprimary $(W,U)$ is given in terms of the generators used in \cite{Shatashvili:1994zw} by
\begin{equation} \label{eq:WUforSpin7}
W=9\Big(
\frac{1}{3}T-X
\Big)\,,
\qquad
U=9\Big(
 \frac{1}{6}G'-M
\Big)\,.
\end{equation}
The operator $T^<$ in \eqref{eq:T<>} happens to be proportional to the operator $X$ which is known  \cite{Shatashvili:1994zw, Figueroa-OFarrill:1996tnk} to be related to the Cayley $4$-form on the $Spin(7)$ target space.\footnote{See e.g.\ \cite{Joyce2007} for an introduction to exceptional holonomy.}

For $q=4$ in the $\nwarrow$ sequence ($\hat{c}=7$), the FS algebra is a subalgebra of the $G_2$ Shatashvili--Vafa algebra \cite{Noyvert:2002mc}; see \eqref{eq:FS_G2}. The superprimary is expressed as
\begin{equation} \label{eq:WUforG2}
W=\frac{27}{2}\Big(\frac{1}{3}T+X
\Big)\,,
\qquad
U=\frac{27}{2}\Big(\frac{1}{6}G'+M
\Big)\,,
\end{equation}
in terms of the generators of \cite{Shatashvili:1994zw}.
$T^<$ is again proportional to $X$, which is again related to a $4$-form on target space, here the $G_2$ co-associative form.

In both algebras, $\hat{c}$ is an integer and matches the dimension of the target. For $q\geq 5$, fractional values of $\hat{c}$ are generically produced in \eqref{eq:nwsequence}, but it turns out, interestingly, that the integer values $\hat{c}=6$, $5$, and  $4$ also arise, respectively for $q=6$, $q=12$, and $q\rightarrow \infty$ (see fig.\,\ref{fig:c^</c_hat}). Also, $q=2$ corresponds to $\hat{c}=10$, the critical dimension of superstrings. These are first hints that $\hat{c}$ may be meaningfully interpreted as a dimension in other cases than $\hat{c}=8$ and $7$. The integers $\hat{c}\in \{0, 1, 2, 3, 4\}$ similarly arise in the $\nearrow$ sequence.

The fact that unitarity allows all these integer values of $\hat{c}$, combined with the known relations with geometry for $\hat{c}=8$ and $7$ suggests that physically reasonable CFTs with chiral symmetry $\mathcal{SW}_{\hat{c}}(\tfrac{3}{2},2)$ may be engineered from non-linear $\sigma$-models with target spaces of dimension $\hat{c}$. The proportionality of $T^<$ to currents $X$ associated to $4$-forms in the $Spin(7)$ and $G_2$ cases further suggests a pivotal role played by $4$-forms. Of course, $4$-forms only exist in integer dimension $\hat{c}$ greater or equal to $4$, which nicely happens to be the accumulation point where the $\nwarrow$ sequence terminates. It is as if the FS algebra $\mathcal{SW}_{\hat{c}}(\tfrac{3}{2},2)$ ``knew'' of $\hat{c}=4$ as a critical minimum dimension. We focus on the $\nwarrow$ sequence for this reason in this note. It is worth mentioning however that  both sequences $\nearrow\,\nwarrow$ share many of the interesting properties listed in the introduction and described in section~\ref{sec:Properties}; see \cite{Gepner:2001px,Naka:2002xs}.
Even if such a direct geometric interpretation does not exist, FS symmetric theories may hold interesting lessons about supersymmetric string backgrounds, in particular non-geometric ones.

In either case, we are led to revisit the properties of Shatashvili--Vafa algebras highlighted in \cite{Shatashvili:1994zw}, aiming to generalize them to the whole FS family. We do this starting in section~\ref{sec:Properties}.
In the next section, we exhibit a third and new link between the FS family and target space geometry via the $\mathcal{N}=2$ superconformal algebra.

\section{Contact with $\mathcal{N}=2$ superconformal algebras} \label{sec:N=2}

Having acknowledged the presence of FS algebras within $\text{SV}^{Spin(7)}$ and $\text{SV}^{G_2}$, we now ask if FS can be found in other chiral algebras known for their importance in supersymmetric string compactifications. The most obvious candidate is the $\mathcal{N}=2$ superconformal algebra, which we denote $\text{\normalfont Vir}^{\mathcal{N}=2}_\textsf{c}$, where $\mathsf{c}$ is the central charge. In the particular cases $\mathsf{c}=3n$, $n\in\mathbb{N}$, this algebra has nice connections
for instance with Calabi--Yau $n$-folds; see e.g.\ \cite{Hori:2003ic, Greene:1996cy, Alim:2012gq, Quigley:2014rya, Neitzke:2004ni, Vonk:2005yv, Wendland:2014ana} for reviews. 

The relevant characteristic of Calabi--Yau $n$-folds here is their $U(n)$-structure.\footnote{Calabi--Yau manifolds of course enjoy a further reduction of structure group $U(n)\rightarrow SU(n)$, and we will address this refinement in section~\ref{sec:Odake}.} They admit a $2$-form---the K\"ahler form---covariantly constant under a connection---Levi--Civita---with holonomy contained in $U(n)$. Perturbatively at least, this yields a worldsheet $U(1)$ chiral symmetry with current $\mathsf{J^3}$ \cite{Howe:1991ic}; see also \cite{delaOssa:2018azc}. $\text{\normalfont Vir}^{\mathcal{N}=2}_\textsf{c}$ is an extension of the $\mathcal{N}=1$ superconformal algebra \eqref{eq:TT}--\eqref{eq:GG}\noeqref{eq:TG} by this current $\mathsf{J^3}$ and its partner $\mathsf{G^3}$, which is a supersymmetry current. Our convention is that $(\mathsf{J^3}, \mathsf{G^3})$ forms a superprimary with respect to the original $\mathcal{N}=1$. The explicit OPEs between all generators are given in appendix~\ref{app:N=2}, where we also connect to the more widespread notation using complexified generators.

We use ``sans serif'' font to distinguish the generators $\mathsf{T}, \mathsf{G}, \mathsf{J^3}, \mathsf{G^3}$ of $\text{\normalfont Vir}^{\mathcal{N}=2}_\textsf{c}$ from the generators $T, G, W, U$ of a putative FS subalgebra, which we want to find inside $\text{\normalfont Vir}^{\mathcal{N}=2}_\textsf{c}$. The central charge $\mathsf{c}$ associated to $\mathsf{T}$ is a priori distinct from the central charge $c$ associated to $T$. We will prove the following.

\vspace{10pt}

\noindent
\textbf{Proposition 1:\quad FS in the $\mathcal{N}=2$ superconformal algebra} \label{prop:1}
\emph{
\begin{enumerate}
\item
Embeddings of the $\mathcal{N}=1$ superconformal algebra $\text{\normalfont Vir}^{\mathcal{N}=1}_c$ inside $\text{\normalfont Vir}^{\mathcal{N}=2}_\mathsf{c}$ exist in a family parametrized by $\theta \in U(1)$ given by
\begin{align}
T &=\mathsf{T}\,, \label{eq:TVir}\\
G &=\cos\theta\,\mathsf{G} +\sin\theta\,\mathsf{G^3}\label{eq:GVir}\,.
\end{align}
In particular, the central charges must match: $c=\mathsf{c}$.
\item 
$\text{\normalfont Vir}^{\mathcal{N}=2}_\mathsf{c}$ contains a unique, up to scale, weight $2$ superprimary with respect to a given $\mathcal{N}=1$ subalgebra (see point 1 above). It is given by the pair
\begin{align}
W&=c_1\Big(\mathsf{T}+\frac{3}{2}\no{\mathsf{J^3J^3}}\Big)\,, \label{eq:WVir}\\
U&=c_1\Big(\cos\theta \big(3\no{\mathsf{G^3J^3}}+2\mathsf{G}'\big)+\sin\theta\big(-3\no{\mathsf{GJ^3}}+2\mathsf{G^3}{}'\big)\Big)\,.
\label{eq:UVir}
\end{align}
It is singular if and only if $\mathsf{c}=1$.
\item
The FS algebra $\mathcal{SW}_{\hat{c}}(\tfrac{3}{2},2)$ with central charge $c=3\hat{c}/2$ embeds inside $\text{\normalfont Vir}^{\mathcal{N}=2}_\mathsf{c}$ with central charge $\mathsf{c}$ if and only if
\begin{equation}
c=\mathsf{c}\text{ equals $6$, $-6$ or $3/2$}\,.\footnote{The embedding is then given by \eqref{eq:TVir}--\eqref{eq:UVir}\noeqref{eq:GVir}\noeqref{eq:WVir} with $c_1$ equal respectively to $9$, $3$ or $-27$.}
\end{equation}
For $c=-6$ and $c=3/2$, the embedding is conditional to quotienting $\text{\normalfont Vir}^{\mathcal{N}=2}_\mathsf{c}$ by singular states.
\end{enumerate}}

\vspace{10pt}

This proposition is the third link \eqref{eq:FS_Vir}, promised in the introduction, between the FS family and geometry. The central charge gets fixed, signalling a link only for $\hat{c}=4$ of all possible positive even target space dimensions. Mysteriously, this is also where the curve on fig.\,\ref{fig:c^</c_hat} reaches its maximum.
The target space geometry has some kind of $U(2)$-structure.

\label{p:VirN=4} K3 surfaces provide the most important example. In this case, the $U(2)$-structure reduces further to $Sp(1)=SU(2)\subset U(2)$ and the K3 surface is hyper-K\"ahler. Such manifolds have a $2$-sphere's worth of K\"ahler $2$-forms, which can be pictured in the $3$-space spanned by three independent K\"ahler $2$-forms $\omega^1, \omega^2, \omega^3$. The chiral algebra is accordingly an $\mathcal{N}=4$ (so-called \emph{small}) superconformal algebra having one $U(1)$ current per K\"ahler form, $\mathsf{J^1}, \mathsf{J^2}, \mathsf{J^3}$, as well as their partners, $\mathsf{G^1}, \mathsf{G^2}, \mathsf{G^3}$, extra supersymmetry currents, in addition to $\mathsf{G}$ and $\mathsf{T}$. Singling out the $\text{\normalfont Vir}^{\mathcal{N}=2}_\mathsf{c=6}$ subalgebra associated to $(\mathsf{J^3},\mathsf{G^3})$ like we did above corresponds to picking a K\"ahler structure. Evidently, proposition 1.1 yields an FS subalgebra associated to any of the possible $\mathcal{N}=2$ subalgebras:
 \begin{equation}
\mathcal{SW}_{4}(\tfrac{3}{2},2) \subset \text{\normalfont Vir}^{\mathcal{N}=2}_{6} \subset \text{\normalfont Vir}^{\mathcal{N}=4}_{6} \,.
\end{equation}

Proposition 1.1 states that there is essentially only the obvious $\mathcal{N}=1$ superconformal algebra inside $\text{\normalfont Vir}^{\mathcal{N}=2}_\mathsf{c}$. The unconstrained angle $\theta$ is a reflection of the $U(1)$ R-symmetry. 
It is easiest to prove the assertion starting with the order $1$ pole in the $GG$ OPE, where for $G$, we take the most general ansatz respecting the desired weight:
\begin{equation}\label{eq:GansatzVir}
G = b_1\mathsf{G}+b_2 \mathsf{G^3}\,.
\end{equation}
We find
\begin{equation}
\wick{\c G(z) \c G(w)}=\cdots + \frac{2(b_1^2+b_2^2)\mathsf{T}(w)}{z-w}\,,
\end{equation}
where we omitted higher order poles. We conclude
$
T=(b_1^2+b_2^2)\mathsf{T}
$. 
Using this, we find
\begin{equation}
\wick{\c T(z) \c T(w)}=\cdots + \frac{(b_1^2+b_2^2)^2\mathsf{T}'(w)}{z-w}\,,
\end{equation}
and conclude $b_1^2+b_2^2=1$. Without loss of generality, $b_1=\cos\theta$ and $b_2=\sin\theta$. This is enough to satisfy all $\text{\normalfont Vir}^{\mathcal{N}=1}_c$ OPEs.

To prove proposition 1.2, we take the general ansatz
\begin{equation}\label{eq:WansatzVir}
W = c_1\mathsf{T}+c_2 \no{\mathsf{J^3J^3}}+c_3 \mathsf{J^3}{}' 
\end{equation}
and compute
\begin{align}
\wick{\c G(z) \c W(w)} &= 
\frac{\big((3c_1/2-c_2)\cos\theta -c_3\sin\theta\big)\mathsf{G}+\big((3c_1/2-c_2)\sin\theta+c_3\cos\theta\big)\mathsf{G^3}}{(z-w)^2}
\\
&\qquad\qquad\qquad\qquad\qquad\qquad\qquad\qquad\qquad\qquad
+ \mathcal{O}((z-w)^{-1})\,.
\end{align}
The order $2$ pole should vanish, which gives $W$ as in \eqref{eq:WVir}. 
The order $1$ pole then determines $U$ exactly as in \eqref{eq:UVir}. It is straightforward to check that the correct OPEs $TW$, $TU$, $GU$ are satisfied.

$W$ is moreover primary with respect to $\mathsf{J^3}$ if an only if, for $m>0$,
\begin{equation}
0=\mathsf{J^3}_{m}W(0)\ket{0}
=\oint \frac{\dd z}{2\pi {{i}}} z^{m} \wick{\c {\mathsf{J^3}}(z) \c W(0)} \ket{0}\,,
\end{equation}
which selects all poles of order greater than $1$. Explicitly,
\begin{equation}
\wick{\c {\mathsf{J^3}}(z) \c W(0)}
=
\frac{c_1(\mathsf{c}-1)\mathsf{J^3}(w)}{(z-w)^2}\,,
\end{equation}
so $\mathsf{c}=1$ is the solution (assuming $c_1\neq 0$). At this central charge, $W(0)\ket{0}$ is automatically singular (annihilated by all positive modes) and thus orthogonal to the whole $\mathcal{N}=2$ superconformal algebra.

Moving on to proposition 1.3, we first rule out the special case $\mathsf{c}=1$ by noticing that the OPE of $W$  with $U$,
\begin{equation}
\wick{\c W(z) \c U(w)} = \frac{-3c_1^2(\mathsf{c}-1)}{(z-w)^3}G(w) + \mathcal{O}((z-w)^{-2})\,, \label{eq:WU_Vir}
\end{equation}
has vanishing order $3$ pole in this case. This is inconsistent with the FS algebra; see \eqref{eq:WU}.  Instead, the leading pole of \eqref{eq:WU_Vir} gives
\begin{equation}
c_1^2(\mathsf{c}-1)=\upmu=(15-\mathsf{c})(4\mathsf{c}+21)\,. \label{eq:c1^2}
\end{equation}
A simple way to curtail the proof is to work at order $2$ in the $WW$ OPE. Comparing with \eqref{eq:WW}, we get the relation \vspace{-1em}
\begin{align}
c_1^2\left( 2\mathsf{T} - 3(\mathsf{c}-2)\no{\mathsf{J^3 J^3}}\right)
&=
2\upmu\mathsf{T}-2\upnu c_1\Big(\mathsf{T}+\frac{3}{2}\no{\mathsf{J^3J^3}}\Big) \,, \\
\Leftrightarrow\quad
2(c_1^2-\upmu+\upnu c_1) \mathsf{T}
+3\big(\upnu c_1-c_1^2(\mathsf{c}-2)\big)\no{\mathsf{J^3J^3}}
&=0 \,, \\
\Leftrightarrow\quad\quad\quad\quad\quad\quad~
\big(c_1(2-\mathsf{c})+\upnu \big) c_1 \big( 2\mathsf{T} +3 \no{\mathsf{J^3J^3}} \big)&=0 \,,
\end{align}
where we used \eqref{eq:c1^2} to reach the third line. The operator appearing here is $W$ and it cannot vanish away from $\mathsf{c}=1$. Hence, we get an equation for the central charge,
\begin{equation}
\pm(2-\mathsf{c})\sqrt{\frac{(15-\mathsf{c})(4\mathsf{c}+21)}{\mathsf{c}-1}}+5\mathsf{c}+6 = 0 \,,
\end{equation}
which is easily solved: $\mathsf{c}=\pm 6$ choosing the plus sign and $\mathsf{c}=3/2$ choosing the minus sign. The corresponding values of $c_1$ follow from \eqref{eq:c1^2}.

The choice $\mathsf{c}=6$ automatically guarantees that all the OPEs agree with FS.

For $\mathsf{c}=-6$ and $c=3/2$, the situation is more complicated as the order $1$ poles $\boxope{WU}_1$ and $\boxope{UU}_1$ do not match. We have however checked that they match up to null fields in the $\mathcal{N}=2$ superconformal algebra. More precisely, we must use that the fields
\begin{align*}
\mathcal{X} &=
(2 \mathsf{c}+7) \mathsf{J^3}{}''-2 \left(2 \mathsf{c}+9\right) \no{\mathsf{TJ^3}}-\left(2 \mathsf{c}+13\right) \no{\mathsf{J^3J^3J^3}}+6 \no{\mathsf{GG^3}} \,,
\\
\mathcal{Y} &= \frac{3 \left(2 \mathsf{c}^2+3 \mathsf{c}-90\right) }{4 \mathsf{c}+21}\mathsf{G}{}''-\frac{36 \left(\mathsf{c}-6\right) }{4 \mathsf{c}+21} \no{\mathsf{TG}}+6 \left(\mathsf{c}-2\right)  \no{\mathsf{G^3J^3}{}'}-4 \left(\mathsf{c}+6\right)  \no{\mathsf{G^3{}'J^3}}+24  \no{\mathsf{GJ^3J^3}} \,,
\\
\widetilde{\mathcal{Y}} &= -3 \left(2 \mathsf{c}+9\right) \mathsf{G^3}{}''+16 \left(\mathsf{c}+6\right)  \no{\mathsf{G'J^3}}+6 \left(2 \mathsf{c}+13\right)  \no{\mathsf{G^3J^3}\mathsf{J^3}}+4 \left(2 \mathsf{c}+3\right)  \no{\mathsf{TG^3}}+12  \no{\mathsf{GJ^3}{}'}
\end{align*}
are singular with respect to $\text{\normalfont Vir}^{\mathcal{N}=2}_\mathsf{c}$ for $\mathsf{c}=-6$ and $\mathsf{c}=3/2$.
The disagreement in $\boxope{WU}_1$ can be expressed as a linear combination of $\mathcal{Y}$ and $\widetilde{\mathcal{Y}}$, while the disagreement in $\boxope{UU}_1$ can be shown to be singular itself.

We do not have an interpretation for the embeddings of the FS algebra at $\mathsf{c}=-6$ and $\mathsf{c}=3/2$. They are a side-result of our analysis which we will not use in the remainder of this paper.

\subsection{$\hat{c}=2n$ and Odake algebras} \label{sec:Odake}

We cannot intuitively explain why the FS algebra fits inside $\text{\normalfont Vir}^{\mathcal{N}=2}_{\mathsf{c}}$ only when $n=\mathsf{c}/3=2$ if we restrict to $n\in\mathbb{N}$. One might have expected FS as well, for example, for $n=1$, $n=3$ and $n=4$, corresponding perhaps to strings on Calabi--Yau $n$-folds. There is however a larger geometrically relevant algebra which might accommodate FS and which we have not considered yet.

Calabi--Yau $n$-folds actually give rise on the worldsheet to \emph{extensions} of $\text{\normalfont Vir}^{\mathcal{N}=2}_{3n}$ by a complex field $\Omega$ of weight $n/2$ and its supersymmetric partner \cite{Odake:1988bh}. We call these extensions, one for each $n\in\mathbb{N}$, \emph{Odake algebras} and denote them $\text{Od}^{n}$. This enhancement of worldsheet chiral symmetry correlates with a structure group reduction of the target space from $U(n)$ to $SU(n)$. The current $\Omega$ is essentially due to the nowhere vanishing holomorphic $n$-form on the Calabi--Yau $n$-fold.

In the $n = 1$ Odake algebra, the extra fields are simply a complex free fermion and its complex $U(1)$ current partner, reflecting the flat geometry of elliptic curves. A realization of the FS algebra in terms of such free fields was actually found in \cite{Gepner:2001px}.
The $n=2$ Odake algebra is identical to the small $\mathcal{N}=4$ superconformal algebra \cite{Odake:1988bh}, and we have proven a contact with K3 surfaces above; see p.\,\pageref{p:VirN=4}.
Inside the $n=4$ Odake algebra, it is also easy to find an FS subalgebra. We can indeed use the embedding
\begin{equation}
\text{SV}^{Spin(7)}\subset \text{Od}^4
\end{equation}
proven in \cite{Figueroa-OFarrill:1996tnk}, interpreted geometrically by the fact that Calabi--Yau $4$-folds in particular support a $Spin(7)$-structure. Using the identification \eqref{eq:FS_Spin7}, we find a $\hat{c}=8$ FS subalgebra inside $\text{Od}^4$. 
Consistently with our proposition 1, this embedding is not fully contained in $\text{Vir}_{\mathsf{c}=12}^{\mathcal{N}=2}\subset \text{Od}^4$, as higher spin fields from $\text{Od}^4$ play a role. We have not searched FS in $\text{Od}^n$ for $n>4$.

So far so good in terms of links with geometry, but the case $n=3$ is more puzzling. We have the following result.

\vspace{10pt}

\noindent
\textbf{Proposition 2:}\quad
Odake's algebra $\text{Od}^{3}$ (central charge $9$) \emph{does not} have any FS subalgebra.

\vspace{10pt}

The proof, similar to the case of the $\mathcal{N}=2$ superconformal algebra, is given in appendix~\ref{app:prop2}. $\text{Od}^3$ superficially looks on track to avoid the no-go result in proposition 1. In addition to the obvious $\mathcal{N}=2$ superconformal subalgebra with $\mathsf{c}=9$, it has another non-commuting $\mathcal{N}=2$ superconformal subalgebra generated by
\begin{equation}
\mathsf{T}^\triangleleft {\;}{{{=}}}{\;} -\frac{1}{6}\no{\mathsf{J^3J^3}} \,,
\qquad
\mathsf{J^3}{}^\triangleleft {\;}{{{=}}}{\;} \frac{\mathsf{J^3}}{3} \,,
\qquad
\mathsf{G}^\triangleleft {\;}{{{=}}}{\;}\frac{{{i}} \mathsf{A}}{\sqrt{6}} \,,
\qquad
\mathsf{G^3}{}^\triangleleft {\;}{{{=}}}{\;}\frac{{{i}} \mathsf{B}}{\sqrt{6}} \,,
\end{equation}
where $\mathsf{A}+{{i}}\mathsf{B}=\Omega$ is the complex field mentioned above. The central charge is $\mathsf{c}^\triangleleft=1$, one of the exceptional values in proposition 1. This subalgebra bears likeness with the hidden small sector $T^<$ of FS algebras. Unfortunately, it falls short of providing even a weight $2$ primary. Indeed, the one given by proposition 1.2 vanishes identically:
\begin{equation}
W^\triangleleft \propto \mathsf{T}^\triangleleft+\frac{3}{2}\no{\mathsf{J^3}{}^\triangleleft\mathsf{J^3}{}^\triangleleft} = 0\,.
\end{equation}

We cannot explain why this case
behaves so differently. The fact remains however that $\text{Od}^3$ and FS at $\hat{c}=6$ have very similar features as we shall discuss in the next section. The comparison between these two algebras yearns for a better understanding.

The contact of FS with the $\mathcal{N}=2$ superconformal algebra at $\hat{c}=4$ is however a very positive result.
It corresponds to $q\rightarrow\infty$ in the unitarity sequence $\nwarrow$ whose properties we now describe.

\section{Unitary representations and special properties} \label{sec:Properties}

Unitary representations of FS algebras were worked out in \cite{Gepner:2001px}. We start here with a recapitulation of the main points.
The FS algebra is like any other $\mathcal{N}=1$ superconformal algebra in that it admits two consistent sets of mode labels: the {Neveu--Schwarz} (NS) and Ramond (R) sectors, where all fermionic modes $G_i, U_i$  are labelled respectively by $i\in \mathbb{Z}+1/2$ or by $i\in\mathbb{Z}$.\footnote{There should be no confusion between the mode label $i$ and the imaginary unit because the FS algebra is over $\mathbb{R}$.} Bosonic modes $L_n, W_n$ are labelled by integers in both sectors, $n\in \mathbb{Z}$. We provide the mode algebras in appendix~\ref{app:FS_ModeAlgebras}.

Representations can be constructed by acting with negative-label modes on highest weight states. Highest weight states are defined to be annihilated by all positive-label modes and they transform in an irreducible representation of the zero mode algebra.
In the NS sector, zero modes arise only from the bosonic generators and they mutually commute. We choose highest weight states as simultaneous eigenvectors and label them by their weights under $T^<$ and $T^>$:
\begin{equation}
L^<_0\ket{h^<,h^>}^{\text{NS}} = h^<\ket{h^<,h^>}^{\text{NS}} \,, \qquad
L^>_0\ket{h^<,h^>}^{\text{NS}} = h^>\ket{h^<,h^>}^{\text{NS}} \,,
\end{equation}
\begin{equation}
\mathcal{O}_{n>0} \ket{h^<,h^>}^{\text{NS}} = 0 \,, \qquad \forall ~ \mathcal{O}\,.
\end{equation}

In the Ramond sector, $L_0$ commutes with all other zero modes, so its eigenvalue (the total weight $h$) can be chosen as label for highest weight states.
The remaining zero modes do not commute. The irreducible representations of the algebra they form (at least when considered on highest weight states) are in general $2$-dimensional. 
However, in the special case where $h=\hat{c}/16$, a $1$-dimensional representation is achieved by taking $G_0=U_0=0$ and $L^<_0=h^<$. 
Such highest weight states are called \emph{Ramond ground states} and we denote them
\begin{equation}
\ket{h^<,\tfrac{\hat{c}}{16}-h^<}^{\text{Rgs}} \,.
\end{equation}

An inner product is introduced via the definition of Hermitian conjugates: 
\begin{equation}
L_n^\dagger {\;}{{{=}}}{\;} L_{-n}\,,\qquad 
G_i^\dagger {\;}{{{=}}}{\;} G_{-i}\,,\qquad 
W_n^\dagger {\;}{{{=}}}{\;} W_{-n}\,,\qquad 
U_i^\dagger {\;}{{{=}}}{\;} -U_{-i}\,.
\end{equation}
Notice the minus sign in $U_n^\dagger$. 
Unitarity imposes in both sectors that the large weight $h^>$ should be non-negative. 
Moreover, $h^<$ should be chosen amongst weights of primaries in the appropriate minimal model, namely \cite{Belavin:1984vu}
\begin{equation} \label{eq:MMweights}
h_{r,s} {\;}{{{=}}}{\;} \frac{\left(r(q+1)-sq\right)^2-1}{4q(q+1)} \,,
\end{equation}
where, for the minimal model labelled by $q\in\{2,3,4,5,\ldots\}$, the integers $r,s$ are restricted to
\begin{equation}
1 \leq r \leq q-1 \,, \qquad 1 \leq s \leq q \,,
\end{equation}
and $(r,s)$ defines the same weight as $(q-r,q+1-s)$. These weights are often pictured in a grid, the \emph{{Kac table}}.
Furthermore, any FS descendant which happens to be primary with respect to $T^<$ must also belong to the Kac table (or be null and quotiented out).
The authors of \cite{Gepner:2001px} analysed these constraints level by level to obtain the FS highest weight states allowed by unitarity.\footnote{They however ignored the cases $q=2$ and $q\rightarrow\infty$, which would require a separate treatment.}

\subsection{Spin field in the hidden small sector} \label{sec:reps}

The Ramond ground states identified in \cite{Gepner:2001px} are reproduced in table~\ref{tab:Rgs}, where we also illustrate a Kac table (for the particular example $q=6$). 
We emphasize that, due to the identification $(r,s)\sim (q-r,q+1-s)$, there are different equivalent ways to picture the states in the Kac table and thus to present the results.
In table~\ref{tab:Rgs}, we give explicitly the two possible ways to write $h^<=h_{r,s} = h_{q-r,q+1-s}$. The first one given is coloured with a darker shade in the Kac table.

\begin{center}
\begin{table}
{
\renewcommand\arraystretch{1.8}
\hskip5pt
\begin{tabular}{c c c c c|c|c|c|c}
    \multicolumn{5}{c|}{$h^<$} &
         $h^>$ &
         $h^<+h^>$ &
         $r,s \in \mathbb{Z}$ & \multirow{3}{*}{        
\put(0,110){ 
\begin{tikzpicture}[transform canvas={scale=0.5}]
\def \q {6}
\draw[->] (0.1,-0.1) -- (1,-0.1) node[right] {\Huge $r$};
\draw[->] (0.1,-0.1) -- (0.1,-1) node[below] {\Huge $s$};
\foreach \r in {1,...,5}{
\foreach \s in {1,...,6}{
\node [draw=black, fill=white, minimum size=1cm] at (\r,-\s) {};
}}
\foreach \r in {1,...,5}{
\node [draw=black, fill=gray, minimum size=1cm] at (\q-\r,-\q+\r-1) {};
\node [draw=black, fill=lightgray, minimum size=1cm] at (\r,-\r) {};
}
\node [draw=black, fill=gray, minimum size=1cm] at (2,-1) {};
\node [draw=black, fill=lightgray, minimum size=1cm] at (\q-2,-\q) {};
\foreach \r in {1,...,5}{
\foreach \s in {1,...,6}{
\pgfmathsetmacro\num{((\r * (\q+1) -\s * \q)^2 -1)/gcd(((\r * (\q+1) -\s * \q)^2 -1),4*\q*(\q+1))}
\pgfmathsetmacro\denom{4*\q*(\q+1)/gcd(((\r * (\q+1) -\s * \q)^2 -1),4*\q*(\q+1))}
\ifdim\denom pt = 1 pt\relax \node at (\r,-\s) {\LARGE  $\pgfmathprintnumber{\num}$};\fi
\ifdim\denom pt > 1 pt\relax \node at (\r,-\s) {\LARGE $\frac{\pgfmathprintnumber{\num}}{\pgfmathprintnumber{\denom}}$};\fi
}}
\end{tikzpicture}
}
         }\\
    \cmidrule[1pt]{1-8}
    \cellcolor{gray} $h_{2,1}$ &
         $=$ &
         \cellcolor{lightgray} $h_{q-2,q}$ &
         $=$ &
         $\tfrac{\hat{c}}{16}$ &
         $0$ &
         $\tfrac{\hat{c}}{16}$ & \\
    \hhline{--------}
    \cellcolor{gray} $h_{q-r,q-r+1}$ &
         $=$ &
         \cellcolor{lightgray} $h_{r,r}$ &  &  & 
         $\frac{(q+2)^2-r^2}{4 q (q+1)}$ &
         $\tfrac{\hat{c}}{16}$ &
         $1 \leq r \leq q-1$ \\
\end{tabular}
}
\caption{Ramond ground states in $\mathcal{SW}_{\hat{c}}(\tfrac{3}{2},2)$ for $\hat{c}=4(q+3)/q$, $q\geq 3$. \\ As a visual guide, we coloured the corresponding cells in the Kac table for $q=6$.}
\label{tab:Rgs}
\end{table}
\end{center}

In the $G_2$ and $Spin(7)$ Shatashvili--Vafa algebras, the hidden minimal model $T^<$ is regarded as providing an analogue of the $U(1)$ sector of $\mathcal{N}=2$ superconformal algebras. A crucial argument for this is the existence of a Ramond ground state 
\begin{equation}
\ket{\tfrac{\hat{c}}{16},0}^{\text{Rgs}} \,,
\end{equation}
sitting \emph{entirely in the small sector}, i.e.\ with $h^>=0$. In the same spirit, the $\mathcal{N}=2$ spin field generating spectral flow by $1/2$,
\begin{equation} \label{eq:Sigma}
\Upsigma {\;}{{{=}}}{\;} \text{exp}\left(-{\frac{{{i}}}{2} \sqrt{\frac{\mathsf{c}}{3}} \mathsf{H}}\right) \,,
\end{equation}
is expressed entirely in terms of the $U(1)$ current $\mathsf{J}={{i}}\sqrt{\frac{\mathsf{c}}{3}}\mathsf{H}'$, where $\mathsf{H}$ is a free boson; see for example \cite{Greene:1996cy}. This is possible because the total weight of Ramond ground states, which is $h={\hat{c}}/{16}$ in any $\mathcal{N}=1$ theory, happens to coincide with Ising and tricritical Ising weights, $1/2$ and $7/16$ respectively, again as if the minimal models ``knew'' about the relevant target space dimensions. The authors of \cite{Shatashvili:1994zw} rate this as ``one of the most remarkable facts for these theories''.

Clearly from table~\ref{tab:Rgs}, this is in no way special to FS algebras with $\hat{c}=8$ or $7$.
\emph{All} members of the $\nwarrow$ sequence have a spin field $\Sigma_{\hat{c}}$ entirely contained in the $T^<$ sector, which creates the Ramond ground state $\ket{h^<,0}^{\text{Rgs}}$ with
\begin{equation} \label{eq:RemarkableH21}
h^<=h_{2,1} = h_{q-2,q} = \frac{q+3}{4q} = \frac{\hat{c}}{16} \,;
\end{equation}
see \eqref{eq:nwsequence} for the last equality.\footnote{A similar statement holds in the $\nearrow$ sequence: there, $h_{1,2}=\hat{c}/16$ is the weight of a Ramond ground state.} 
This ground state allows to predict most of the remarkable properties of FS theories, including topological twisting (see section~\ref{sec:Twists}), at least to some degree, and a discrete notion of spectral flow (see section~\ref{sec:ChiralAntichiralNSPrimaries}).

\subsubsection{Spacetime supersymmetry} \label{sec:SpacetimeSUSY}

$\Sigma_{\hat{c}}$ is also crucial for the RNS construction of supercharges \cite{Friedan:1985ey, Friedan:1985ge} for $(10-\hat{c})$-dimensional spacetime similar to the construction of \cite{Banks:1987cy} using $\Upsigma$, $\Upsigma^\dagger$ and the $\mathcal{N}=2$ superconformal algebra. At least this is true for heterotic strings when $\hat{c}=8$ \cite{Melnikov:2017yvz}. 
The case $\hat{c}=7$ works similarly: the authors of \cite{Melnikov:2017yvz} proved the necessity of the $\text{SV}^{G_2}$ algebra on the worldsheet for spacetime supersymmetry. It is natural to ask however if the enhancement from $\mathcal{SW}_{7}(\tfrac{3}{2},2)$ to $\text{SV}^{G_2}$ really plays a important role or if, perhaps, the FS subalgebra could be sufficient.

It would also be interesting to revisit the original work \cite{Banks:1987cy} in $\hat{c}=6$ to see what spacetime consequence, if not 4d supersymmetry, has the FS algebra $\mathcal{SW}_{6}(\tfrac{3}{2},2)$. Spacetime supersymmetry may also be a useful approach to find a meaning or application to FS at $\hat{c}=5$. We have not pursued these ideas very far.

\subsection{Spectral flow to special NS primaries} \label{sec:ChiralAntichiralNSPrimaries}

In CFTs with chiral $\mathcal{N}=2$ superconformal symmetry, the spectral flow operator \eqref{eq:Sigma} realizes a bijection between Ramond ground states and certain NS states, called \emph{chiral} or \emph{anti-chiral primaries} depending on the direction of the flow \cite{Lerche:1989uy}.

In FS theories, analogous states can be predicted using $\Sigma_{\hat{c}}$ and fusion.
The Ramond spin field $\Sigma_{\hat{c}}$ being fully contained in the small sector, it has particularly simple fusion rules dictated by the minimal model. Let us introduce notation for highest weight states with respect to $T^<$, which also have definite external weight:
\begin{align}
L^<_0 \{h^<,h^>\} &= h^<\{h^<,h^>\} \,, \qquad L^>_0 \{h^<,h^>\} = h^>\{h^<,h^>\} \,, \\
L^<_{n} \{h^<,h^>\} &= 0\,, \qquad n>0 \,.
\end{align}
By standard minimal model fusion rules, we can write
\begin{equation} \label{eq:fusionSigmahatc}
\ket{h_{2,1}=\tfrac{\hat{c}}{16},0}^{\text{Rgs}} \times \{h_{r,s},h^>\} = \{h_{r-1,s}, h^>\} + \{h_{r+1,s}, h^>\} \,,
\end{equation}
(unless $r$ is $1$ or $q-1$, in which case one of the terms on the right hand side is absent).

For example, fusion of $\ket{\tfrac{\hat{c}}{16},0}^{\text{Rgs}}$ with itself generates $\{0,0\}$ and, provided  $q\geq 4$, also the state $\{h_{3,1},0\}$.
Identifying the former with the vacuum, we notice a further similarity with the $\mathcal{N}=2$ case, where the spin field $\Upsigma$ also maps to the vacuum upon spectral flow by $-1/2$. Flowing by $+1/2$, $\Upsigma$ maps to $\text{exp}\left(-{{{i}} \sqrt{\frac{\mathsf{c}}{3}} \mathsf{H}}\right)$ which is then somewhat similar to $\{h_{3,1},0\}$.\footnote{We should say that we suspect the distinction between $\Upsigma$ and $\Upsigma^\dagger$ to be irrelevant in
this rough fusion argument, and we would like to recall that $\Upsigma^\dagger$ maps to $\text{exp}\left(+{{{i}} \sqrt{\frac{\mathsf{c}}{3}} \mathsf{H}}\right)$ upon flow by $-1/2$ and to the vacuum upon flow by $+1/2$.}

Applied to all FS Ramond ground states, this fusion argument shines light on finitely many states which turn out to exist in the discrete spectrum of NS highest weight states found in \cite{Gepner:2001px}. This is straightforward to verify by comparing with appendix B of this reference.\footnote{There appears to be a typo in \cite{Gepner:2001px}. In the last row of the first table in section B.2, we should read $2\leq n$.} We will call these states \emph{special}. They are listed in table~\ref{tab:SpecialNSstates}, where we also coloured cells in a sample Kac table ($q=6$) to communicate the states visually more clearly. We coloured in shades of red the image of table~\ref{tab:Rgs} translated to the left, i.e.\ $r\rightarrow r-1$, and in shades of blue the result of a translation to the right. Paler shades are copies of darker shades of the other colour, so there is no reason at this stage to distinguish two colours, but this will be useful later.

The remark that certain NS states in FS theories share characterisitics with $\mathcal{N}=2$ chiral primaries was first made in \cite{Shatashvili:1994zw} specifically in the cases $\hat{c}=8$ and $7$. The idea was further studied in \cite{deBoer:2005pt} in the case $\hat{c}=7$.

In the limit $q\rightarrow\infty$, corresponding to $\hat{c}=4$, we should expect from section~\ref{sec:N=2} consistency with the $\mathcal{N}=2$ results. We can make a simple observation going along these lines: the ratio of FS Ramond ground states to special NS states approaches 1:2, consistently with the fact that there is half as many Ramond ground states in an $\mathcal{N}=2$ CFT as there are chiral/anti-chiral states.
For finite $q\geq 4$, there are $q$ FS Ramond ground states and $2(q-1)$ special NS states (i.e.\ $q-1$ cells in the Kac table~\ref{tab:SpecialNSstates} of a given colour). $q=3$ is a boundary case with only three special NS states.

\begin{table}
\begin{center}
{
\renewcommand\arraystretch{1.8}
\hskip-70pt
\begin{tabular}{c c c c c|c|c|c|c}
    \multicolumn{5}{c|}{$h^<$} &
         $h^>$ &
         $h^<+h^>$ &
         $r \in \mathbb{Z}$ & \multirow{5}{*}{
\put(0,80){
\begin{tikzpicture}[transform canvas={scale=0.5}]
\def \q {6}
\draw[->] (0.1,-0.1) -- (1,-0.1) node[right] {\Huge $r$};
\draw[->] (0.1,-0.1) -- (0.1,-1) node[below] {\Huge $s$};
\foreach \r in {1,...,5}{
\foreach \s in {1,...,6}{
\node [draw=black, fill=white, minimum size=1cm] at (\r,-\s) {};
}}
\foreach \r in {2,...,5}{
\node [draw=black, fill=MyPaleRed, minimum size=1cm] at (\r-1,-\r) {};
}
\foreach \r in {1,...,4}{
\node [draw=black, fill=MyRed, minimum size=1cm] at (\q-\r-1,-\q+\r-1) {};
}
\node [draw=black, fill=MyRed, minimum size=1cm] at (1,-1) {};
\node [draw=black, fill=MyPaleRed, minimum size=1cm] at (\q-3,-\q) {};
\foreach \r in {1,...,4}{
\node [draw=black, fill=MyPaleBlue, minimum size=1cm] at (\r+1,-\r) {};
}
\foreach \r in {2,...,5}{
\node [draw=black, fill=MyBlue, minimum size=1cm] at (\q-\r+1,-\q+\r-1) {};
}
\node [draw=black, fill=MyBlue, minimum size=1cm] at (3,-1) {};
\node [draw=black, fill=MyPaleBlue, minimum size=1cm] at (\q-1,-\q) {};
\foreach \r in {1,...,5}{
\foreach \s in {1,...,6}{
\pgfmathsetmacro\num{((\r * (\q+1) -\s * \q)^2 -1)/gcd(((\r * (\q+1) -\s * \q)^2 -1),4*\q*(\q+1))}
\pgfmathsetmacro\denom{4*\q*(\q+1)/gcd(((\r * (\q+1) -\s * \q)^2 -1),4*\q*(\q+1))}
\ifdim\denom pt = 1 pt\relax \node at (\r,-\s) {\LARGE  $\pgfmathprintnumber{\num}$};\fi
\ifdim\denom pt > 1 pt\relax \node at (\r,-\s) {\LARGE $\frac{\pgfmathprintnumber{\num}}{\pgfmathprintnumber{\denom}}$};\fi
}}
\end{tikzpicture}
}
         }\\
    \cmidrule[1pt]{1-8}
    \cellcolor{MyRed} $h_{1,1}$ &
         $=$ &
         \cellcolor{MyPaleBlue} $h_{q-1,q}$ &
         $=$ &
         $0$ &
         $0$ &
         $0$ & \\
    \hhline{--------}
    \cellcolor{MyRed} $h_{q-r-1,q-r+1}$ &
         $=$ &
         \cellcolor{MyPaleBlue} $h_{r+1,r}$ &  &  &
         $\frac{(q+2)^2-r^2}{4 q (q+1)}$ &
         $\frac{q+r+2}{2q}$ &
         $1 \leq r \leq q-2$ \\
    \hhline{--------}
    \cellcolor{MyBlue} $h_{3,1}$ &
         $=$ &
         \cellcolor{MyPaleRed} $h_{q-3,q}$ &  &  &
         $0$ &
         $\frac{q+2}{q}$ & \\
    \hhline{--------}
    \cellcolor{MyBlue} $h_{q-r+1,q-r+1}$ &
         $=$ &
         \cellcolor{MyPaleRed} $h_{r-1,r}$ &  &  & 
         $\frac{(q+2)^2-r^2}{4 q (q+1)}$ &
         $\frac{q-r+2}{2q}$ &
         $2 \leq r \leq q-1$ \\
\end{tabular}
}
\caption{Special Neveu--Schwarz primaries in $\mathcal{SW}_{\hat{c}}(\tfrac{3}{2},2)$ for $\hat{c}=4(q+3)/q$, $q\geq 3$. (The third line concerns only $q\geq 4$.) As a visual guide, we also coloured the corresponding cells in the Kac table for $q=6$.}
\label{tab:SpecialNSstates}
\end{center}
\end{table}

\subsubsection{Unitarity bound} \label{sec:UnitarityBound}

Special primaries saturate a unitarity bound analogous to the $\mathcal{N}=2$ BPS bound $|\mathsf{q}|/2 \leq \mathsf{h}$.\footnote{$\mathsf{h}$ and $\mathsf{q}$ are respectively eigenvalues of the zero modes of $\mathsf{T}$ and $\mathsf{J}$ in the conventions of appendix~\ref{app:N=2}.} This was noticed within the  Shatashvili--Vafa algebra $\text{SV}^{G_2}$ in \cite{deBoer:2005pt}, but we explain below that the bound actually follows from the FS subalgebra $\mathcal{SW}_{7}(\tfrac{3}{2},2)\subset \text{SV}^{G_2}$. This fact simplifies the proof given in \cite{deBoer:2005pt}, essentially because the subalgebra is insensitive to redundancies introduced by the null ideal of $\text{SV}^{G_2}$ \cite{Figueroa-OFarrill:1996tnk}. The general result, valid for any $q$, can actually be extracted from the analysis in \cite{Gepner:2001px} of vanishing curves of the FS Kac determinant ($d_1^{\text{NS}}$, eq.\,7.10 in that reference). We propose the following simple derivation.

Let us consider, in the NS sector of an FS theory, the level $1/2$ descendants: $G_{-1/2}\ket{h^<, h^>}^{\text{NS}}$ and $U_{-1/2}\ket{h^<, h^>}^{\text{NS}}$. Computing their inner products with the commutation relations, we find the Kac matrix
\begin{equation}
\mathcal{M}{\;}{{{=}}}{\;}
\left[
\begin{matrix}
2h & 2w \\
2w & 
2 \left((c+12) w-54 (c-15) h^2+h \left(2 c^2-33 c-54 w+45\right)\right)
\end{matrix}
\right]\,,
\end{equation}
where $w$ is the eigenvalue of $W_0$. Its determinant is
\begin{equation}
\det \mathcal{M}=
36 (c+12) h^< \big(18 h^2-(c-6) h-(c+12) h^<\big)\,,
\end{equation}
where we have assumed $c>-12$ and traded $w$ for the hidden small weight. This determinant should be non-negative in a unitary theory. Given also that $h^<$ is non-negative, we immediately get
\begin{equation} \label{eq:unitaritybound}
h^< \leq \frac{h (18h - c + 6)}{c+12}\,.
\end{equation}

The bound reduces to that of \cite{deBoer:2005pt} if we substitute $\hat{c}=7$. For $\hat{c}=8$, it applies to the $Spin(7)$ Shatashvili--Vafa algebra, a fact which is perhaps not easily appreciated from the current literature.

As we mentioned, \eqref{eq:unitaritybound} is saturated by special primaries for all $q\geq 3$. This is easy to check with the data in table~\ref{tab:SpecialNSstates}. We did not verify if others amongst the list of NS states of \cite{Gepner:2001px} also saturate the bound, but this could easily be answered for any fixed value of $q$. Representations built on special primaries are therefore short. An eigenvector of $\mathcal{M}$ has vanishing eigenvalue. It is null and should be quotiented out. Explicitly, we find
\begin{equation} \label{eq:SpecialShortening}
\left((3-2c+54h)G_{-1/2}+U_{-1/2}\right)\ket{h^<,h^>}^{\text{special NS}}\sim 0 \,.
\end{equation}

For $\hat{c}=4$, the bound \eqref{eq:unitaritybound} becomes $h^<\leq h^2$, which is not immediately recognizable in the context of $\mathcal{N}=2$ CFT.
The situation clarifies if we invoke the Virasoro operator $\mathsf{T}^\triangleleft$ constructible from the $U(1)$ current. For $\mathsf{J}$ defined like in appendix~\ref{app:N=2}, we have
\begin{equation} \label{eq:Ttriangle}
\mathsf{T}^\triangleleft {\;}{{{=}}}{\;} \frac{3}{2\mathsf{c}}\no{\mathsf{JJ}}\,.
\end{equation}
This operator generalizes the one presented in section~\ref{sec:Odake} in the context of Odake algebras and the operator $T^<$ obtained in $\mathcal{SW}_{4}(\tfrac{3}{2},2) \subset \text{\normalfont Vir}^{\mathcal{N}=2}_{6}$ by proposition 1 (p.\,\pageref{prop:1}).

We have
\begin{equation}
\mathsf{L}^\triangleleft_0 = \frac{3}{2\mathsf{c}}\Big(\mathsf{J_0J_0}+
\sum_{m=1}^\infty \mathsf{J}_{-m}\mathsf{J}_{m}
\Big)\,,
\end{equation}
where the infinite sum is positive semi-definite.  Thus, for any state with well-defined charge $\mathsf{q}$ and weight $\mathsf{h}^\triangleleft$ under $\mathsf{T}^\triangleleft$, we have the first inequality below
\begin{equation}
\mathsf{h}^\triangleleft \geq \frac{3}{2\mathsf{c}}\mathsf{q}^2 \leq \frac{6}{\mathsf{c}}\mathsf{h}^2\,,
\end{equation}
while the second one is the BPS bound. The first turns into an equality if and only if the state is primary with respect to $\mathsf{J}$. We then have an exact match with the FS bound provided $\mathsf{c}=6$.

In particular, chiral and anti-chiral $\mathcal{N}=2$ primaries saturate the FS bound at this central charge.
This is suggestive of a reorganization of FS special states into $\mathcal{N}=2$ chiral/anti-chiral states in the limit $q\rightarrow\infty$ ($\hat{c}=4$). Unfortunately, we cannot draw definitive conclusions at this time, because of the lack of a proper analysis of NS primaries, special or not, in unitary FS theories at $\hat{c}=4$.

\label{p:Ring} A natural question is whether the FS special primaries close a non-singular ring under the OPE. This is normally argued from the BPS bound in $\mathcal{N}=2$ theories, so we may hope for a similar application of the FS bound \eqref{eq:unitaritybound}, see also \cite{Shatashvili:1994zw, deBoer:2005pt}.
An important difference is that the eigenvalue of $T^<$ is not additive, unlike charge with respect to $\mathsf{J}$. Minimal model fusion may however offer compensative help.

\subsubsection{A candidate for marginal deformations} \label{sec:MarginalDef}

We briefly point out that the special highest weight state with $h^<=h_{1,2}$ (last row of table~\ref{tab:SpecialNSstates}) is the only one with total dimension $1/2$, and this is true \emph{for all $q$} thanks to the relation
\begin{equation} \label{eq:magicrelation}
h_{1,2}+h_{2,1}-h_{2,2} = \frac{1}{2} \qquad \forall q\,.
\end{equation}

In the $G_2$ and $Spin(7)$ cases, this state is known to be related to marginal deformations of the CFT. More precisely, in type II context, restoring the anti-holomorphic degrees of freedom, it has been proven  that
\begin{equation}
G_{-1/2}\ket{h_{1,2},\tfrac{1}{2}-h_{1,2}}^{\text{NS}}\otimes \overline{G_{-1/2}\ket{h_{1,2},\tfrac{1}{2}-h_{1,2}}^{\text{NS}}}
\end{equation}is exactly marginal to all orders in perturbation theory and that it also preserves the Ising (tricritical Ising) symmetry. As a matter of fact, this last statement follows directly from the shortening condition \eqref{eq:SpecialShortening} valid for FS special primaries. This is easy to see by combining with \eqref{eq:WUforSpin7} or \eqref{eq:WUforG2} and comparing with the null states found in \cite{Shatashvili:1994zw} (their equations 3.23 and 3.50).

Along a similar direction, we mention the work \cite{Fiset:2017auc}, where constraints on deformations of heterotic $G_2$ compactifications were obtained from a $\sigma$-model argument built on $\text{SV}^{G_2}$.
Inspection reveals that the FS subalgebra $\mathcal{SW}_{7}(\tfrac{3}{2},2)\subset \text{SV}^{G_2}$ is sufficient for the argument in that paper to work.

It is probable that generalizations exist to all unitary FS CFTs. If this were true, this would be another similarity with $\mathcal{N}=2$ CFTs, where marginal deformations are well-known to be constructed from a certain subset of chiral and anti-chiral primaries.
We leave the full analysis to future work, but one quick check confirming this intuition is the proof that the minimal model symmetry is preserved. Following \cite{Shatashvili:1994zw}, we compute
\begin{align} \label{eq:MarginalityMinimalSector}
L^<_0G_{-1/2}&\ket{h_{1,2},\tfrac{1}{2}-h_{1,2}}^{\text{NS}}
\\
&=
\bigg(
h_{1,2}G_{-1/2}
+\frac{c^<}{c}[L_0,G_{-1/2}]-\frac{1}{3(c+12)}[W_0,G_{-1/2}]
\bigg)
\ket{h_{1,2},\tfrac{1}{2}-h_{1,2}}^{\text{NS}}\,,
\end{align}
and we will show that this is proportional to the null descendant \eqref{eq:SpecialShortening} of $\ket{h_{1,2},\tfrac{1}{2}-h_{1,2}}^{\text{NS}}$, just like in Shatashvili--Vafa algebras. We will conclude that the candidate marginal deformation acts trivially in the minimal model sector, as it should.

It is useful to note the remarkable relation
\begin{equation} \label{eq:RemarkableH12}
h_{1,2}=h_{q-1,q-1}=\frac{q-2}{4(q+1)}=\frac{c^<}{\hat{c}}\,,
\end{equation}
which is to be compared with \eqref{eq:RemarkableH21}:
\begin{equation}
h_{2,1} = h_{q-2,q} = \frac{q+3}{4q} = \frac{\hat{c}}{16} \,.
\end{equation}
Thus, $h_{1,2}$ has the simple interpretation as the \emph{slope} of the straight line relating the origin to a given point in the $\nwarrow$ sequence in figure~\ref{fig:c^</c_hat}. The situation exactly dualizes in the $\nearrow$ sequence, where we have instead $h_{1,2}={\hat{c}}/{16}$ and $h_{2,1}={c^<}/{\hat{c}}$.

Returning to \eqref{eq:MarginalityMinimalSector}, we get
\begin{align}
L^<_0G_{-1/2}&\ket{h_{1,2},\tfrac{1}{2}-h_{1,2}}^{\text{NS}}\\
&=
\bigg(
h_{1,2}G_{-1/2}
+\frac{c^<}{2c}G_{-1/2}+\frac{1}{3(c+12)}U_{-1/2}
\bigg)
\ket{h_{1,2},\tfrac{1}{2}-h_{1,2}}^{\text{NS}} \\
&=
\frac{1}{3(c+12)}
\Big(2(15-c)
G_{-1/2}
+
U_{-1/2}
\Big)
\ket{h_{1,2},\tfrac{1}{2}-h_{1,2}}^{\text{NS}}\,,
\end{align}
where we expressed $h_{1,2}$ in terms of $h=1/2$ using that the unitarity bound \eqref{eq:unitaritybound} is saturated. This is now manifestly proportional to \eqref{eq:SpecialShortening}.

\subsection{$G$ is $\lbrace h_{1,2} ,3/2-h_{1,2} \rbrace$}

An important feature of FS theories is that the supersymmetry current $G$ is not only primary with respect to $T$ (with weight $3/2$), but also primary with respect to $T^<$. We indeed have
\begin{align}
\wick{\c {T^<}(z) \c G(w)} 
&=\frac{c^<}{c}\wick{\c T(z) \c G(w)}-\frac{1}{\sqrt{\nu^2+4\mu}}\wick{\c W(z) \c G(w)} \\
&=\frac{c^<}{\hat{c}}\frac{G(w)}{(z-w)^2}+\frac{1}{z-w}\left(\frac{c^<}{c}G'(w)+\frac{U(w)}{\sqrt{\nu^2+4\mu}}\right)\,,
\end{align}
from which we can read off the small weight of $G$. In the $\nwarrow$ sequence, $G$ creates the state $\{h_{1,2},\frac{3}{2}-h_{1,2}\}$ for all $q$, in the notation of section~\ref{sec:ChiralAntichiralNSPrimaries}. We used \eqref{eq:RemarkableH12}. In fact the FS algebra itself, for any fixed value of $q$, may be \emph{constructed} \cite{Gepner:2001px} as the extension of the $q$-minimal model by its primary of weight $h_{1,2}$.

The $(r,s)=(1,2)$ primary has particularly simple fusion rules, similar to \eqref{eq:fusionSigmahatc}. Focusing on the small sector, we have
\begin{equation} \label{eq:fusionPhi12}
\{h_{1,2}\}\times\{h_{r,s}\}=\{h_{r,s-1}\}+\{h_{r,s+1}\}\,.
\end{equation}
This yields the following reinterpretation of the result in the previous section about marginal deformations. Applying $G_{-1/2}$ on $\ket{h_{1,2},\tfrac{1}{2}-h_{1,2}}^{\text{NS}}$ produces states with values of $h^<$  immediately above and below $h_{1,2}$ in the Kac table, i.e.\ $h^<=0$ and $h^<=h_{1,3}$ (or descendants thereof). Acting further with $L_0^<$ gave a null state, which we interpret to mean that only the $h^<=0$ conformal family has actually been produced. In other words, $\ket{h_{1,2},\tfrac{1}{2}-h_{1,2}}^{\text{NS}}$ is in the kernel of the part of $G_{-1/2}$ mapping to the $(r,s)=(1,3)$ conformal family. This should be reminiscent again of $\mathcal{N}=(2,2)$ CFTs, where marginal deformations are also in the kernel of some operators constructed from supersymmetry charges. These operators play another role as the BRST charges of topologically twisted theories. We elaborate in the next part.

\section{Topological twists} \label{sec:Twists}

After revisiting the role of the spin field \eqref{eq:Sigma} in topological twists of $\mathcal{N}=(2,2)$ CFTs, Shatashvili and Vafa \cite{Shatashvili:1994zw} argue, replacing it by $\Sigma_{\hat{c}}$ (see section~\ref{sec:reps}), for the existence of topological twists of $\sigma$-models with $G_2$ or $Spin(7)$ targets. This part, relying heavily on the Coulomb gas representation of Virasoro minimal models, is by and large conjectural, and subtelties were highlighted in \cite{deBoer:2005pt}. Our aim here is to go over the main arguments in \cite{Shatashvili:1994zw} and show that they generalize nicely to the whole $\nwarrow$ sequence of FS algebras. Building on \cite{deBoer:2005pt}, we will then attempt to reach a refined interpretation. The context in this part is type II string theory, but most arguments apply chiraly, so we shall suppress anti-holomorphic degrees of freedom. The $\nearrow$ sequence should also behave similarly.

Twisting $\mathcal{N}=(2,2)$ SCFTs \cite{Witten:1988xj, Witten:1988ze, Witten:1991zz} is achieved by adding background $U(1)$ gauge fields, which amounts to a redefinition of covariant derivatives. Effectively, this modifies the 2d action by
\begin{equation} \label{eq:deltaStwist}
\delta S =\int \dd^2 \sigma  \, \frac{1}{2}\left(
\mp \mathsf{J}\,\omega_{\bar{z}} + \overline{\mathsf{J}}\,\omega_z
\right)\,,
\end{equation}
where $\mathsf{J}$ and $\overline{\mathsf{J}}$ are the holomorphic and anti-holomorphic $U(1)$ currents and $\omega$ is the spin connection. 
The holomorphic energy-momentum tensor in particular gets redefined as
\begin{equation} \label{eq:N=2TwistedVirasoro}
\mathsf{T}_{\text{twisted}}{\;}{{{=}}}{\;}
\mathsf{T}\pm\frac{1}{2} \partial_{z}\mathsf{J}\,.
\end{equation}
The $(\pm)$ choice leads to the difference between the A and B twists.

Bosonizing the $U(1)$ current reveals a formulation \cite{Bershadsky:1993cx, Antoniadis:1993ze} more hopeful to define twists of $\mathcal{N}=(1,1)$ CFTs. Using $\mathsf{H}$ in \eqref{eq:Sigma}, we have
\begin{equation}
\int \dd^2\sigma \,  \frac{1}{2}\mathsf{J}\,\omega_{\bar{z}}
= \int \dd^2\sigma  \, \frac{{{i}}}{2} \sqrt{\frac{\mathsf{c}}{3}} \partial_{z}\mathsf{H}\,\omega_{\bar{z}}
= -\int \dd^2\sigma  \, \frac{{{i}}}{2} \sqrt{\frac{\mathsf{c}}{3}} \mathsf{H}\,\partial_{z}\omega_{\bar{z}}\,.
\end{equation}
Here $\partial_{z}\omega_{\bar{z}}$ is essentially the Ricci scalar, which, on a genus $g$ surface, can be chosen to have delta-function support at $2-2g$ points. 
On the sphere for example, twisting can then be regarded as the insertion of two spin fields\footnote{We were not too careful with signs, but choosing a different sign in \eqref{eq:N=2TwistedVirasoro} amounts to replacing $\Upsigma$ by $\Upsigma^\dagger$ in \eqref{eq:twistCorrelators}.} at infinity:
\begin{equation} \label{eq:twistCorrelators}
\langle \ldots \rangle^{S^2}_\text{twisted} {\;}{{{=}}}{\;} \langle \ldots (\Upsigma(\infty))^2\rangle^{S^2} \,.
\end{equation}

The central charge of both twisted Virasoro operators \eqref{eq:N=2TwistedVirasoro} is zero. Moreover, chiral fields have dimension zero under the $(+)$-twisted Virasoro operator, while anti-chiral fields have dimension zero under the $(-)$-twisted Virasoro operator.

We will shortly observe similar properties in FS theories through the lens of the Coulomb gas representation \cite{Feigin:1981st, Felder:1988zp}---see e.g. \cite{DiFrancesco:1997nk,Mussardo:2010mgq} for introductions. We will now review this framework briefly, focusing on the relevant formul\ae. The main idea is to bosonize the minimal model sector using a free holomorphic boson $H$ (which plays the role of $\mathsf{H}$ in $\mathcal{N}=2$ theories). We take the logarithmic OPE
\begin{equation}
\wick{ \c H(z) \c H(w)} = -\log(z-w).
\end{equation}
Before twisting, the minimal model stress-tensor is written as
\begin{equation} \label{eq:TCoulomb}
T^<=\frac{1}{2}\no{jj}+\alpha_0  j' \,,\qquad\text{where}\qquad j = {{i}} H '\,.
\end{equation}
The central charge of \eqref{eq:TCoulomb} is
\begin{equation} \label{eq:c_alpha0Coulomb}
c^<=1-12\left(\alpha_0\right)^2 \, \quad \text{and thus } \quad \alpha_0 = \frac{1}{\sqrt{2q(q+1)}} \,.
\end{equation}

Primaries under $T^<$ are represented by exponentials,
\begin{equation} \label{eq:CoulombV}
V_\alpha(z){\;}{{{=}}}{\;}e^{{{i}}\alpha H} \,.
\end{equation}
$V_{\alpha}$ has weight $\frac{1}{2}\alpha (\alpha - 2\alpha_0)$ with respect to $T^<$. In particular, both $V_\alpha$ and $V_{2\alpha_0-\alpha}$ describe the same primary. This ambiguity translates in the identification $(r,s) \sim (q-r,q+1-s)$ of minimal model primaries mentioned in section~\ref{sec:Properties}. The charge of $V_\alpha$ with respect to $j$ is $\alpha$. Charges of minimal model primaries are
\begin{equation} \label{eq:CoulombCharges}
\alpha_{r,s}{\;}{{{=}}}{\;}\frac{1}{\sqrt{2}}\left[(1-r)\sqrt{\frac{q+1}{q}}-(1-s)\sqrt{\frac{q}{q+1}}\right] \,,
\end{equation}
and the corresponding weights are as given by \eqref{eq:MMweights}:
\begin{equation}
h_{r,s}=\frac{1}{2}\alpha_{r,s}(\alpha_{r,s}-2\alpha_0)=\frac{\left(r(q+1)-sq\right)^2-1}{4q(q+1)} \,.
\end{equation}

In the Coulomb gas representation, the minimal model is viewed as a deformation parametrized by $\alpha_0$ of the ordinary free bosonic theory. This is especially manifest in the expression \eqref{eq:TCoulomb} for the Virasoro operator. By the same token as for the $U(1)$ sector of $\mathcal{N}=2$ theories, this deformation can be thought of as the insertion of vertex operators
in free boson correlation functions; see \eqref{eq:twistCorrelators}. The concrete upshot is that a twist defined by \eqref{eq:twistCorrelators} manifests itself as a \emph{shift in $\alpha_0$}.

Either $(r,s)=(2,1)$ or $(r,s)=(q-2,q)$ can be used to represent the Ramond spin field $\Sigma_{\hat{c}}$ in FS theories, hence there appears to be two possible ways to shift $\alpha_0$, and thus two possible twists. We call them \emph{$(+)$-twist} and \emph{$(-)$-twist}. Only the $(+)$-twist features in the existing literature on $G_2$ and $Spin(7)$ CFTs. Perhaps this is for good reasons---it has somewhat better properties than the $(-)$-twist. We mention both because of analogies with the $\mathcal{N}=2$ case. The twisted background charges are
\begin{align} \label{eq:Atwist}
{\alpha_0}_{(+)\text{-twisted}}
&{\;}{{{=}}}{\;} \alpha_0 - \alpha_{2,1} ~~~ = +\frac{q+2}{\sqrt{2q(q+1)}}\,, \\
{\alpha_0}_{(-)\text{-twisted}}
&{\;}{{{=}}}{\;} \alpha_0 - \alpha_{q-2,q} =-\frac{q+2}{\sqrt{2q(q+1)}}\,.
\label{eq:Btwist}
\end{align}

Note that these differ only by a sign. Moreover for $q\rightarrow \infty$, they become $\pm 1/\sqrt{2}$ which allows to make direct contact with the $\mathcal{N}=2$ $(\pm)$-twists. Identifying $H = \mathsf{H}$ in the limit, the twisted stress-tensors indeed become
\begin{align}
{T^<}_{(\pm)\text{-twisted}}
&\overset{q\rightarrow\infty}{\longrightarrow} \frac{1}{2} \no{jj} \pm \frac{1}{\sqrt{2}} j' \\
&~=~ \frac{3}{2\mathsf{c}} \no{\mathsf{JJ}} \pm \frac{{{i}}}{\sqrt{2}}\mathsf{H}''\\
&~=~ \mathsf{T^\triangleleft} \pm 
\sqrt{\frac{3}{2\mathsf{c}}}\mathsf{J}'\,,
\end{align}
where we used the definition of $\mathsf{J}$ near \eqref{eq:Sigma} and  $\mathsf{T}^\triangleleft$ in \eqref{eq:Ttriangle}. This matches exactly \eqref{eq:N=2TwistedVirasoro} provided $\mathsf{c}=6$, explaining in particular our choice of the names ``$(\pm)$-twists'' in FS theories.

\subsection{Vanishing central charge} \label{sec:VanishingCentralCharge}

Recall that the total central charge of FS theories splits as $c = c^< + c^>$. The twists \eqref{eq:Atwist} and \eqref{eq:Btwist} only affect the internal sector, so $c^>$ remains unchanged. However the small central charge becomes
\begin{equation}
{c^<}_{(\pm)\text{-twisted}}{\;}{{{=}}}{\;}1-12\big({\alpha_0}_{(\pm)\text{-twisted}}\big)^2 \,,
\end{equation}
see \eqref{eq:c_alpha0Coulomb}. It is a simple exercise to check that \emph{this is exactly $-c^>$ for all values of $q$} for both twists. We have then obtained what Shatashvili and Vafa call the ``strongest hint for the existence of a topological theory'' \cite{Shatashvili:1994zw}: 
After either $(+)$ or $(-)$ twist, the total central charge of the system vanishes,
\begin{equation}
{c^<}_{(\pm)\text{-twisted}} + c^> = 0 \,.
\end{equation}
A zero central charge is a smoking gun of topological theories.

\subsection{Dimensionless special states} \label{sec:dimensionlesschirals}

Let us now consider what happens upon twisting to the total weight of the NS states that we called ``special''.
Large weights $h^>$ remain unchanged. Small weights $h^<$ after twist are computed with the redefined $\alpha_0$:
\begin{equation}
{h^<_{r,s}}_{(\pm)\text{-twisted}} {\;}{{{=}}}{\;} \frac{1}{2}\alpha_{r,s}\big(\alpha_{r,s}-2{\alpha_0}_{(\pm)\text{-twisted}}\big) \,.
\end{equation}

Note that this breaks the symmetry about the center of the Kac table, which means that weights computed after twist depend on the Coulomb gas representation of primaries chosen before the twist. As an example, take $q=3$ and consider the $(+)$-twist. The vacuum represented as $V_{\alpha_{1,1}}$ maps to $\ket{0,0}$, while representing it as $V_{\alpha_{q-1,q}}$ yields $\ket{-1/3,0}$. We will return to this issue, but the interesting point in \cite{Shatashvili:1994zw} that we presently seek to reproduce is that, \emph{for all special primaries, one of the two representations has vanishing total dimension after $(+)$-twist}.

The effect of the $(+)$-twist is shown in table~\ref{tab:SpecialNSstates_Atwisted}. The Coulomb gas representations in red and pink in the Kac table (corresponding to a ``translation to the left'') are dimensionless after $(+)$-twist. They are in some sense the ``chiral'' ones; incidentally, they have $j$-charge larger or equal to zero, just like $\mathcal{N}=2$ chiral primaries. Complementary representations---dark and pale blue cells---remain dimensionful. We may call them ``anti-chiral'': they have negative $j$-charge (except $\alpha_{q-1,q}=\sqrt{2}/\sqrt{q(q+1)}>0$).

\begin{table}
\begin{center}
{
\renewcommand\arraystretch{1.8}
\hskip-70pt
\begin{tabular}{c c|c c|c|c}
    \multicolumn{4}{c|}{${h^<_{r,s}}_{(+)\text{-twisted}}+h^>$} &
         $r \in \mathbb{Z}$ & \multirow{5}{*}{
\put(0,80){
\begin{tikzpicture}[transform canvas={scale=0.5}]
\def \q {6}
\draw[->] (0.1,-0.1) -- (1,-0.1) node[right] {\Huge $r$};
\draw[->] (0.1,-0.1) -- (0.1,-1) node[below] {\Huge $s$};
\foreach \r in {1,...,5}{
\foreach \s in {1,...,6}{
\node [draw=black, fill=white, minimum size=1cm] at (\r,-\s) {};
}}
\foreach \r in {2,...,5}{
\node [draw=black, fill=MyPaleRed, minimum size=1cm] at (\r-1,-\r) {\LARGE $0$};
}
\foreach \r in {1,...,4}{
\node [draw=black, fill=MyRed, minimum size=1cm] at (\q-\r-1,-\q+\r-1) {\LARGE $0$};
}
\node [draw=black, fill=MyRed, minimum size=1cm] at (1,-1) {\LARGE $0$};
\node [draw=black, fill=MyPaleRed, minimum size=1cm] at (\q-3,-\q) {\LARGE $0$};
\foreach \r in {1,...,4}{
}
\foreach \r in {2,...,5}{
}
\end{tikzpicture}
}
         }\\
    \cmidrule[1pt]{1-5}
    \cellcolor{MyRed} $(1,1):$ &
         \cellcolor{MyRed} $0$ &
         $(q-1,q):$ &
         $-\frac{1}{q}$ & \\
    \hhline{-----}
    \cellcolor{MyRed} $(q-r-1,q-r+1):$ &
         \cellcolor{MyRed} $0$ &
         $(r+1,r):$ &
         $\frac{q+r+1}{q}$  &
         $1 \leq r \leq q-2$ \\
    \hhline{-----}
    $(3,1):$ &
         $\frac{2q+3}{q}$ &
         \cellcolor{MyPaleRed} $(q-3,q):$ &
         \cellcolor{MyPaleRed} $0$ &\\
    \hhline{-----}
    $(q-r+1,q-r+1):$ &
         $\frac{q-r+1}{q}$ &
         \cellcolor{MyPaleRed} $(r-1,r):$ &
         \cellcolor{MyPaleRed} $0$ &
         $2 \leq r \leq q-1$ \\
\end{tabular}
}
\caption{$(+)$-twist on special Neveu--Schwarz states. The numbers $(r,s)$ in parenthesis determine the Coulomb gas representation used to calculate the weights after twist. One of the two possible representations (red and pink cells) is dimensionless after $(+)$-twist.}
\label{tab:SpecialNSstates_Atwisted}
\end{center}
\end{table}

The $(-)$-twist displays a complementary behaviour, see table~\ref{tab:SpecialNSstates_Btwisted}, although only in the limit $q\rightarrow\infty$. Most of the special primaries are not dimensionless after $(-)$-twist for finite $q$. In the limit $q\rightarrow\infty$ however, the Coulomb representations having non-vanishing $(+)$-twisted weight (blue cells) have vanishing $(-)$-twisted weight and vice versa. An exception is $V_{\alpha_{1,1}}$ which is dimensionless after either twist. Both representations of the vacuum are in fact dimensionless as $q\rightarrow \infty$ after either twist. The field $V_{\alpha_{2,1}}$ is also remarkable because its $(-)$-twisted dimension is zero even for finite $q$.

\begin{table}
\begin{center}
{
\renewcommand\arraystretch{1.8}
\hskip-65pt
\begin{tabular}{c c|c c|c|c}
    \multicolumn{4}{c|}{${h^<_{r,s}}_{(-)\text{-twisted}}+h^>$} &
         $r \in \mathbb{Z}$ & \multirow{5}{*}{
\put(0,80){
\begin{tikzpicture}[transform canvas={scale=0.5}]
\def \q {6}
\draw[->] (0.1,-0.1) -- (1,-0.1) node[right] {\Huge $r$};
\draw[->] (0.1,-0.1) -- (0.1,-1) node[below] {\Huge $s$};
\foreach \r in {1,...,5}{
\foreach \s in {1,...,6}{
\node [draw=black, fill=white, minimum size=1cm] at (\r,-\s) {};
}}
\foreach \r in {2,...,5}{
}
\foreach \r in {1,...,4}{
}
\foreach \r in {1,...,4}{
\node [draw=black, fill=MyPaleBlue, minimum size=1cm] at (\r+1,-\r) {};
}
\foreach \r in {2,...,5}{
\node [draw=black, fill=MyBlue, minimum size=1cm] at (\q-\r+1,-\q+\r-1) {};
}
\node [draw=black, fill=MyBlue, minimum size=1cm] at (3,-1) {};
\node [draw=black, fill=MyPaleBlue, minimum size=1cm] at (\q-1,-\q) {};
\end{tikzpicture}
}
         }\\
    \cmidrule[1pt]{1-5}
    $(1,1):$ &
         $0$ &
         \cellcolor{MyPaleBlue} $(q-1,q):$ &
         \cellcolor{MyPaleBlue} $\frac{q+3}{q(q+1)}$ & \\
    \hhline{-----}
    $(q-r-1,q-r+1):$ &
         $\frac{(q+2)(q+r+2)}{q(q+1)}$ &
         \cellcolor{MyPaleBlue} $(r+1,r):$ &
         \cellcolor{MyPaleBlue} $\frac{1-r}{q(q+1)}$  &
         $1 \leq r \leq q-2$ \\
    \hhline{-----}
    \cellcolor{MyBlue}
    $(3,1):$ &
         \cellcolor{MyBlue}
         $-\frac{1}{q}$ &
         $(q-3,q):$ &
         $\frac{2(q+2)^2}{q(q+1)}$ &\\
    \hhline{-----}
    \cellcolor{MyBlue}
    $(q-r+1,q-r+1):$ &
         \cellcolor{MyBlue}
         $\frac{1+r}{q(q+1)}$ &
         $(r-1,r):$ &
         $\frac{(q+2)(q-r+2)}{q(q+1)}$ &
         $2 \leq r \leq q-1$ \\
\end{tabular}
}
\caption{$(-)$-twist on special Neveu--Schwarz states. The numbers $(r,s)$ in parenthesis determine the Coulomb gas representation used to calculate the weights after twist. One of the two possible representations (dark and pale blue cells) is dimensionless after $(-)$-twist in the limit $q\rightarrow\infty$.}
\label{tab:SpecialNSstates_Btwisted}
\end{center}
\end{table}

\subsection{Conformal blocks and BRST operator}

The arguments we have developed so far in this section, while highly suggestive of topologically twisted theories, are subtle to make rigourous. The Coulomb gas representation really introduces new degrees of freedom absent from the original theory. $T^<$ is part of the theory, but not $j$ on its own, so charge is meaningless, unlike in $\mathcal{N}=2$ CFTs. The boson $H$ is also unphysical, as are the various vertex operator representations of primaries. One can also dress vertex operators with multiple screening charges; changing their charge, but preserving how they transform under the conformal group. Rigourously, the free boson Fock space must be restricted as the cohomology of some BRST operator constructed from some screened vertex operator. We refer to \cite{DiFrancesco:1997nk, Felder:1988zp} for details.

Because of these complications, the authors of \cite{deBoer:2005pt} attempted to define the $(+)$-twist in $\text{SV}^{G_2}$ theories independently of the Coulomb gas representation. Doing so relied heavily on fusion of $T^<$ conformal families and on their conformal block decomposition. As an important example, consider again the fusion rules \eqref{eq:fusionSigmahatc} and \eqref{eq:fusionPhi12}:
\begin{align}
\{h_{2,1}\}\times\{h_{r,s}\}&=\{h_{r-1,s}\}+\{h_{r+1,s}\}\,, \\
\{h_{1,2}\}\times\{h_{r,s}\}&=\{h_{r,s-1}\}+\{h_{r,s+1}\}\,.
\end{align}
Projecting the right hand side to either of the neighbouring families of $(r,s)$ in the Kac table defines conformal block decompositions
\begin{align}
\{h_{2,1}\} &= \{h_{2,1}\}^\leftarrow + \{h_{2,1}\}^\rightarrow \,, \\
\{h_{1,2}\} &= \,\{h_{1,2}\}^\uparrow \, + \,\{h_{1,2}\}^\downarrow \,.
\end{align}
The arrows convey the idea of moving left/right or up/down in the Kac table.

The authors of \cite{deBoer:2005pt} propose to define the twist by inserting $\Sigma_{\hat{c}}^\rightarrow$ in correlation functions, as opposed to $\Sigma_{\hat{c}}$ as described at the beginning of this section. 
One should think of the left/right decomposition of the Ramond spin field $\Sigma_{\hat{c}}$ as essentially analogous to the distinction between the $\mathcal{N}=2$ spin fields $\Upsigma^\dagger$ and $\Upsigma$, which generate spectral flow by $-1/2$ and $+1/2$ respectively.
The new contact with $\mathcal{N}=2$ we provided in section~\ref{sec:N=2} supports this interpretation. This is strikingly illustrated in the Coulomb gas framework:
\begin{align}
\Sigma_{\hat{c}}^\rightarrow ~ \sim ~\,~ V_{\alpha_{2,1}}~ &\overset{q\rightarrow\infty}{\longrightarrow} ~ \text{exp}\left(-\frac{{{i}}}{\sqrt{2}}\mathsf{H}\right) = \Upsigma \,,
\\
\Sigma_{\hat{c}}^\leftarrow ~ \sim ~ V_{\alpha_{q-2,q}} &\overset{q\rightarrow\infty}{\longrightarrow} ~ \text{exp}\left(+\frac{{{i}}}{\sqrt{2}}\mathsf{H}\right) = \Upsigma^\dagger
\,,
\end{align}
where we used \eqref{eq:CoulombV}, \eqref{eq:CoulombCharges} and \eqref{eq:Sigma}.

We believe the reason why the arguments in sections~\ref{sec:VanishingCentralCharge}, \ref{sec:dimensionlesschirals} are so compelling can be traced to the relationships between screened vertex operators in the Coulomb gas approach and conformal blocks of minimal model primaries. (This is the meaning of ``$\sim$'' above.) By distinguishing Coulomb representations of primaries, we were roughly accessing individual blocks. By \eqref{eq:Btwist}, we were noting that an insertion of $\Sigma_{\hat{c}}^\leftarrow$ might also be worth considering. Insertion of $\Sigma_{\hat{c}}^{\rightarrow}$ or $\Sigma_{\hat{c}}^{\leftarrow}$ should affect in different ways the conformal blocks of a given primary, mirroring how the charge of different Coulomb representations of primaries changes differently upon \eqref{eq:Atwist} or \eqref{eq:Btwist}. We refer the reader to \cite{deBoer:2005pt} for more insightful comments. The details deserve to be clarified, and we hope to return to this in a future publication.

Another important proposal of \cite{deBoer:2005pt} is a BRST operator controlling the twisted theory. It is made of down conformal blocks of the holomorphic and anti-holomorphic supersymmetry charges. Restraining ourselves to the holomorphic sector, we are talking about $G^\downarrow_{-1/2}$. The identification is again strongly suggested by the contact with $\mathcal{N}=2$ and the Coulomb formalism:
\begin{align}
G^\downarrow ~ \sim ~~~ V_{\alpha_{1,2}}\widetilde{G} ~~~~ &\overset{q\rightarrow\infty}{\longrightarrow} ~ \text{exp}\left(+\frac{{{i}}}{\sqrt{2}}\mathsf{H}\right)\widetilde{G} = \mathsf{G}^+ \,,\\
G^\uparrow ~ \sim ~ V_{\alpha_{q-1,q-1}}\widetilde{G} ~ &\overset{q\rightarrow\infty}{\longrightarrow} ~ \text{exp}\left(-\frac{{{i}}}{\sqrt{2}}\mathsf{H}\right)\widetilde{G} = \mathsf{G}^-\,.
\end{align}
Compare for instance with \cite[eq.~3.42]{Greene:1996cy}. Here $\widetilde{G}$ is the part of $G$ which is primary with respect to $T-T^<$ with weight $\tfrac{3}{2}-h_{1,2}$. All the charge with respect to $\mathsf{J}={{i}}\sqrt{2}\mathsf{H}'$ is carried by the exponential so this reproduces the expected $\mathsf{J}$-charge $\pm 1$ of $\mathsf{G}^\pm$. It is also straightforward to evaluate the Coulomb prediction for twisted weights:
\begin{align}
{h^<_{1,2}}_{(+)\text{-twisted}}+\frac{3}{2}-h_{1,2} &= 1\,,\qquad\qquad\quad~~
{h^<_{q-1,q-1}}_{(+)\text{-twisted}}+\frac{3}{2}-h_{1,2} = \frac{q-1}{q}+1\,,
\\
{h^<_{1,2}}_{(-)\text{-twisted}}+\frac{3}{2}-h_{1,2} &= \frac{q+2}{q+1}+1\,,\qquad
{h^<_{q-1,q-1}}_{(-)\text{-twisted}}+\frac{3}{2}-h_{1,2} = \frac{3}{q(q+1)}+1\,.
\end{align}
Consistently with the evidence presented in \cite{Shatashvili:1994zw}, $V_{\alpha_{1,2}}$ is weight $1$ after $(+)$-twist, which we would expect of a BRST current. Again this is actually true for all $q$. Meanwhile $V_{\alpha_{q-1,q-1}}$  becomes dimension $1$ after $(-)$-twist albeit only in the limit $q\rightarrow\infty$.

To substantiate their BRST operator, the authors of \cite{deBoer:2005pt} point out that $(G^\downarrow_{-1/2})^2 = (G^\uparrow_{-1/2})^2=0$ by virtue of the $\mathcal{N}=1$ anti-commutation relations. It proved however difficult in \cite{deBoer:2005pt} to formulate a BRST-exact twisted stress-tensor, although they had promising results for certain modes, such as $L_{-1}=\{G^\downarrow_{-1/2},G^\uparrow_{-1/2}\}$. A twisted stress-tensor defined via conformal blocks would improve upon the arguments in section~\ref{sec:VanishingCentralCharge}. We trust that the new contact with $\mathcal{N}=2$ we have uncovered will help inspire a solution to this problem.

A related open question is the cohomology of $G^\downarrow_{-1/2}$ and $G^\uparrow_{-1/2}$. Recall that chiral (resp.\ anti-chiral) $\mathcal{N}=2$  primaries are annihilated by $\mathsf{G}^+$ (resp.\ $(\mathsf{G}^+)^*=\mathsf{G}^-$). This is in fact their very definition. Similarly, one expects some conformal blocks of special primaries to be annihilated by $G^\downarrow_{-1/2}$ or $G^\uparrow_{-1/2}$. There are results along those line for the $G_2$ case in \cite{deBoer:2005pt}. This is presumably the improved interpretation of ``chiral'' (red) and ``anti-chiral'' (blue) Coulomb representations we had in section~\ref{sec:dimensionlesschirals}.
While we cannot easily test this at the level of blocks, we argue below in this direction in the language of states, borrowing from \cite{Gepner:2001px}. This will be a generalization of the computation made at the end of section \ref{sec:MarginalDef} and an improvement upon section 4.4.1 of \cite{deBoer:2005pt}.\vspace{10pt}

Let $\ket{h^<, h-h^<}^{\text{NS}}$ be a Neveu--Schwarz highest weight state with respect to the FS algebra for a fixed value of $q$. 
Level $1/2$ descendants are automatically annihilated by positive modes of $T^<$:
\begin{equation}
L^<_{n}G_{-1/2}\ket{h^<, h-h^<}^{\text{NS}}
=
L^<_{n}U_{-1/2}\ket{h^<, h-h^<}^{\text{NS}}
=
0\,,
\qquad
n>0 \,.
\end{equation}
It is moreover possible to construct two linear combinations which are eigenstates of $L^<_0$ and therefore highest weight with respect to $T^<$. In our notation from section~\ref{sec:ChiralAntichiralNSPrimaries}, they are
\begin{align}
\{\check{h}^<, h+\tfrac{1}{2}-\check{h}^<\}
\qquad \text{and} \qquad
\{\hat{h}^<, h+\tfrac{1}{2}-\hat{h}^<\}\,,
\end{align}
where the small weights are related to $h^<$ and $c$ through
\begin{align}
\check{h} &= h^<+\frac{9+\sqrt{(c-6)^2+72h^<(c+12)}}{2(c+12)}\,,\\
\hat{h} &= h^<+\frac{9-\sqrt{(c-6)^2+72h^<(c+12)}}{2(c+12)}\,.
\end{align}

We again focus on the $\nwarrow$ sequence, so we express $c$ in terms of $q$ using \eqref{eq:nwsequence}.
Choosing also $h^<=h_{r,s}$, we find
\begin{equation}
\check{h}= 
\begin{cases}
h_{r,s+1}  & \text{if } sq \geq r(q+1) \,, \\
h_{r,s-1}  & \text{otherwise,}
\end{cases}
\qquad \text{and} \qquad
\hat{h}=
\begin{cases}
h_{r,s-1} & \text{if } sq \geq r(q+1)\,, \\
h_{r,s+1} & \text{otherwise.}
\end{cases}
\end{equation}
This gives a handy characterization of the up/down decomposition of $G_{-1/2}$, at least on highest weight states. We find
\begin{align}
G^\downarrow_{-1/2}\ket{h^<,h^>}^{\text{NS}}
&=
\left(
\frac{1}{2}G_{-1/2}
+\frac{\left(2U_{-1/2}-(c+12)G_{-1/2}\right)}{6\sqrt{(c-6)^2+72h^<(c+12)}}
\right)
\ket{h^<, h^>}^{\text{NS}}
=
\{\check{h}^<, h+\tfrac{1}{2}-\check{h}^<\}\,,
\\
G^\uparrow_{-1/2}\ket{h^<,h^>}^{\text{NS}}
&=
\left(
\frac{1}{2}G_{-1/2}
-\frac{\left(2U_{-1/2}-(c+12)G_{-1/2}\right)}{6\sqrt{(c-6)^2+72h^<(c+12)}}
\right)
\ket{h^<, h^>}^{\text{NS}}
=
\{\hat{h}^<, h+\tfrac{1}{2}-\hat{h}^<\}\,.
\end{align}
The denominator never vanishes in the unitary regime except for the very specific case $c=6$ and $h^<=0$.

Let us now generalize a result obtained at the end of section \ref{sec:Properties} by restricting to special NS primaries. We can use that they saturate the unitarity bound \eqref{eq:unitaritybound} to find that the square root appearing above becomes $\sqrt{(c-6-36h)^2}$. Except for the case $h=0$, which must be ignored, inspection of table~\ref{tab:SpecialNSstates} shows that all special primaries have $h=h^<+h^>$ greater than $(c-6)/36=1/2q$, so the minus sign must be chosen. Simplifying further gives
\begin{equation}
G^\downarrow_{-1/2}\ket{h^<,h^>}^{\text{special NS}}
=
-\frac{\left(
(3-2c+54h)G_{-1/2}
+
U_{-1/2}
\right)
\ket{h^<, h^>}^{\text{special NS}}}{3(c-6-36h)}
=
0\,,
\end{equation}
which is exactly the level $1/2$ null state identified in section~\ref{sec:UnitarityBound}. Special primaries are thus $G^\downarrow_{-1/2}$-closed.

We can also check that $G^\downarrow_{-1/2}\ket{h^<,h^>}^{\text{NS}}$ reduces to $\mathsf{G}^\pm\ket{h^<,h^>}^{\text{NS}}$ when $\hat{c}=4$ under appropriate circumstances. For simplicity we take $\theta=0$ in the embedding $\mathcal{SW}_4(\tfrac{3}{2},2)\subset \text{Vir}^{\mathcal{N}=2}_6$, so that
\begin{align*}
U_{-1/2}
&= 18 (\mathsf{G}')_{-1/2}+27\no{\mathsf{G^3J^3}}_{-1/2} \\
&= -18 \mathsf{G}_{-1/2} + 
27 \left(\sum_{m\leq -3/2} \mathsf{G^3}_m \mathsf{J^3}_{-1/2-m}
+
\sum_{m\geq -1/2} \mathsf{J^3}_{-1/2-m} \mathsf{G^3}_m \right) \\
&= 9 \mathsf{G}_{-1/2} + 27\mathsf{G^3}_{-1/2}\mathsf{J^3}_{0} + 
27\left(\sum_{m\leq -3/2} \mathsf{G^3}_m \mathsf{J^3}_{-1/2-m}
+
\sum_{m\geq 1/2} \mathsf{J^3}_{-1/2-m} \mathsf{G^3}_m\right) \,.
\end{align*}
If we assume $\ket{h^<,h^>}^{\text{NS}}$ is primary with respect to $\mathsf{J}=-{{i}}\mathsf{J^3}$ with charge $\mathsf{q}$, then we know from section \ref{sec:UnitarityBound} that $\mathsf{q}^2=4h^<$. Moreover the sum in parenthesis in the line above vanishes. In this case,
\begin{align*}
G^\downarrow_{-1/2}\ket{h^<,h^>}^{\text{NS}}
&=
\left(
\frac{1}{2}G_{-1/2}
+\frac{\left(2U_{-1/2}-18G_{-1/2}\right)}{6\sqrt{72\cdot 18 h^<}}
\right)
\ket{h^<, h^>}^{\text{NS}} \\
&=
\frac{1}{2}
\left(
\mathsf{G}_{-1/2}
+\frac{{{i}}\mathsf{q}}{2\sqrt{h^<}}\mathsf{G^3}_{-1/2}
\right)
\ket{h^<, h^>}^{\text{NS}} \\
&=\frac{1}{\sqrt{2}}\mathsf{G}^\pm_{-1/2}\ket{h^<, h^>}^{\text{NS}} \,.
\end{align*}

\section{Conclusion} \label{sec:Discussion}

We highlighted in this paper the FS subsymmetry $\mathcal{SW}_{\hat{c}}(\tfrac{3}{2},2)$ governing most of the interesting features of conformal 2d $\sigma$-models with either 7d $G_2$ or 8d $Spin(7)$ target space. Proving its presence as a subalgebra of the $\mathcal{N}=2$ superconformal algebra if and only if $\hat{c}=\pm 4$ or $1$, we also explained the source of the many analogies with $\mathcal{N}=2$ CFTs noticed in the past. The case $\hat{c}=4$ further provides a new link between FS and geometry, in particular K3 surfaces. We also suggested applications of FS in supersymmetric string backgrounds for various target space dimensions $\hat{c} \in \{1, 2, 3, 4, 5, 6, 7, 8,~ 10\}$, all of which admit unitary representations.

Unitary FS theories are labelled by their Virasoro minimal model subsymmetry. They all contain a spin field entirely in the minimal model sector, whose left/right fusion rules define Neveu--Schwarz primaries that we called \emph{special}. We argued that the latter are analogous to $\mathcal{N}=2$ chiral/anti-chiral primaries in many respect. We described the unitarity bound they saturate and compared to the $\mathcal{N}=2$ BPS bound at $\hat{c}=4$. One particular special primary looks promising to construct an exactly marginal deformation, again for all unitary FS theories (although we focused on the $\nwarrow$ sequence for definiteness). 

We finally speculated on ($\pm$) topological twists. We generalized the evidence presented in \cite{Shatashvili:1994zw}, based on the Coulomb gas representation of minimal models, to all unitary FS theories ($\nwarrow$ sequence). Of the two apparent possibilities, the $(+)$-twist is the most promising: twisted dimension zero for one of the two Coulomb representations of special primaries. However, both seem to admit a twisted Virasoro operator with vanishing central charge. They also lead to the usual A and B twists for $\hat{c}=4$. 

It is not clear how to twist without the Coulomb gas formalism, but it seems block decompositions of minimal model primaries are an important element. We presented evidence that the splitting $G=G^{\uparrow}+G^{\downarrow}$ is analogous to the splitting between $\mathsf{G}^+$ and $\mathsf{G}^-$ in $\mathcal{N}=2$ CFTs. Similarly, the blocks $\Sigma_{\hat{c}}^\leftarrow$ and $\Sigma_{\hat{c}}^\rightarrow$ of the remarkable FS spin field seem analogous to the $\mathcal{N}=2$ spin fields $\Upsigma$ and $\Upsigma^\dagger$.

Open research directions include the following.
\begin{enumerate}
\item Revisit the construction of spacetime supercharges from $\hat{c}$-dimensional internal CFTs to hopefully explain the role or meaning of unitary FS theories at $\hat c=5$, $6$, $7$ and $10$ in particular, see section~\ref{sec:SpacetimeSUSY}.
\item Examine directly unitary representations of the FS algebra for $\hat{c}=10$ ($q=2$) and $\hat{c}=4$ ($q\rightarrow\infty$). The latter is especially important given the contact with $\mathcal{N}=2$ and its potential to clarify applications of FS theories in general. 
\item Decide if an analogue of the chiral ring is formed by special primaries in general FS theories, see p.\,\pageref{p:Ring}.
\item Clarify the conformal block decomposition and define twists more rigorously. A BRST-exact energy-momentum tensor is an important missing element. Again, the contact we provided with the $\mathcal{N}=2$ case should advise on the route to take.
\item Characterize more systematically cohomology classes of the BRST operator $G^\downarrow_{-1/2}$. Links with certain geometric cohomology of the target space are expected, like in the Calabi--Yau case \cite{deBoer:2005pt, Fiset:2019ecu}.
\item Define new protected observables or topological invariants for $G_2$ and $Spin(7)$ $\sigma$-models similarly to what has been done for Calabi--Yau manifolds.
\end{enumerate}

\section*{Acknowledgements}

For discussions related to this project, I wish to acknowledge Xenia de la Ossa, Matthias Gaberdiel, Christopher Beem, Katrin Wendland, Sebastian Goette, Johannes Walcher, Ingmar Saberi, Sakura Sch\"afer-Nameki, Ruben Minasian, Ilarion Melnikov, Andreas Braun and Michele Del Zotto. I also benefited from Kris Thielemans’ OPEdef Mathematica package.
My work is supported by a grant of the Swiss National Science Foundation.
This research was also partly supported by the NCCR SwissMAP, funded by the Swiss National Science Foundation, and by the Mathematical Institute at the University of Oxford.

\newpage

\appendix

\section{$\mathcal{N}=2$ superconformal algebra} \label{app:N=2}

Generators:
\begin{itemize}
\item $\mathsf{T}$ : real bosonic weight $2$ Virasoro operator;
\item $\mathsf{G}$ : real fermionic weight $3/2$ supersymmetry current;
\item $\mathsf{J^3}$ : real bosonic weight $1$ $U(1)$ current;
\item $\mathsf{G^3}$ : real fermionic weight $3/2$ supersymmetry current.
\end{itemize}

Singular OPEs:
\begin{align}
\wick{\c {\mathsf{T}}(z) \c {\mathsf{T}}(w)}&=
\frac{\mathsf{c}/2}{(z-w)^4}+\frac{2{\mathsf{T}}(w)}{(z-w)^2}+\frac{{\mathsf{T}}'(w)}{z-w}\,,\\
\wick{\c {\mathsf{T}}(z) \c {\mathsf{G}}(w)} &=
\frac{3/2}{(z-w)^2}{\mathsf{G}}(w)+\frac{{\mathsf{G}}'(w)}{z-w}\,,\\
\wick{\c {\mathsf{G}}(z) \c {\mathsf{G}}(w)} &=
\frac{2\mathsf{c}/3}{(z-w)^3}+\frac{2{\mathsf{T}}(w)}{z-w}\,,\\
\wick{\c {\mathsf{T}}(z) \c {\mathsf{J^3}}(w)} &=
\frac{{\mathsf{J^3}}(w)}{(z-w)^2}+\frac{{\mathsf{J^3}}{}'(w)}{z-w}\,,\\
\wick{\c {\mathsf{T}}(z)  \c {\mathsf{G^3}}(w)} &=
\frac{3/2}{(z-w)^2}{\mathsf{G^3}}(w)+\frac{\mathsf{G^3}{}'(w)}{z-w}\,,\\
\wick{\c {\mathsf{G}}(z) \c {\mathsf{J^3}}(w)} &=
\frac{{\mathsf{G^3}}(w)}{z-w}\,,\\
\wick{\c {\mathsf{G}}(z) \c {\mathsf{G^3}}(w)} &=
\frac{2{\mathsf{J^3}}(w)}{(z-w)^2}+\frac{{\mathsf{J^3}}{}'(w)}{z-w}\,,\\
\wick{\c {\mathsf{G^3}}(z) \c {\mathsf{G^3}}(w)} &=
\frac{2\mathsf{c}/3}{(z-w)^3}+\frac{2{\mathsf{T}}(w)}{z-w}\,,\\
\wick{\c {\mathsf{J^3}}(z) \c {\mathsf{J^3}}(w)} &=
-\frac{\mathsf{c}/3}{(z-w)^2}\,,\\
\wick{\c {\mathsf{G^3}}(z) \c {\mathsf{J^3}}(w)} &=
-\frac{{\mathsf{G}}(w)}{z-w}\,.
\end{align}

A different basis is common in the literature.
We define
\begin{equation}
\mathsf{J}{\;}{{{=}}}{\;}-{{i}}\mathsf{J^3} \,, \qquad\qquad\qquad
\mathsf{G^\pm}{\;}{{{=}}}{\;}\frac{1}{\sqrt{2}}\left(\mathsf{G} \pm {{i}}\mathsf{G^3}\right).
\end{equation}
$\mathsf{G^+}$ and $\mathsf{G^-}$ are complex conjugate: $\mathsf{G^-}=(\mathsf{G^+})^*$. The OPEs not involving $\mathsf{T}$ become:
\begin{align}
\wick{\c {\mathsf{J}}(z) \c {\mathsf{J}}(w)}
&=\frac{\mathsf{c}/3}{(z-w)^2} \,, \\
\wick{\c {\mathsf{J}}(z) \c {\mathsf{G^\pm}}(w)}
&=\pm\frac{\mathsf{G^\pm}(w)}{z-w} \,, \\
\wick{\c {\mathsf{G^+}}(z) \c {\mathsf{G^-}}(w)}
&=\frac{2\mathsf{c}/3}{(z-w)^3}+\frac{2{\mathsf{J}}(w)}{(z-w)^2}+\frac{({\mathsf{J}}'+2\mathsf{T})(w)}{z-w} \,, \\
\wick{\c {\mathsf{G^\pm}}(z) \c {\mathsf{G^\pm}}(w)}
&=0 \,.
\end{align}

\section{Proof of inexistence of $\mathcal{SW}_{\hat{c}}(\frac{3}{2},2)$ in $\text{Od}^3$} \label{app:prop2}

The approach is the same as for proposition 1. We start with the most general ansatz for FS generators (italic letters) allowed by dimensional analysis in the Odake algebra (sans serif letters). Our Odake generators and OPEs are as in \cite{Fiset:2018huv}.
\begin{align}
T &= a_1\mathsf{T}+a_2 \no{\mathsf{J^3J^3}}+a_3 \mathsf{J^3}{}' +a_4 \mathsf{C}+a_5 \mathsf{D}\,, \label{eqTansatzOd3}\\
G &= b_1\mathsf{G}+b_2 \mathsf{G^3} + b_3 \mathsf{A} + b_4\mathsf{B}\,, 
\label{eq:GansatzOd3}\\
W &= c_1\mathsf{T}+c_2 \no{\mathsf{J^3J^3}}+c_3 \mathsf{J^3}{}' + c_4 \mathsf{C}+c_5 \mathsf{D}\,,
\label{eq:WansatzOd3}
\\
U &= d_1\mathsf{G}{}'+d_2\mathsf{G^3}{}'+d_3\no{\mathsf{GJ^3}}+d_4\no{\mathsf{G^3J^3}} + d_5 \mathsf{A}' + d_6 \mathsf{B}'\,.
\label{eq:UansatzOd3}
\end{align}

The next step is to impose FS OPEs to fix the coefficients. It is useful to start with
\begin{align}
\wick{\c G(z) \c G(w)} &= \frac{6 (b_1^2 + b_2^2) - 4 (b_3^2 + b_4^2)}{(z-w)^3}\\
&\qquad+
\frac{2\big(
(b_1^2+b_2^2)\mathsf{T}
+(b_3^2 + b_4^2)\no{\mathsf{J^3J^3}}
+(b_1 b_3 + b_2 b_4)\mathsf{C}
+(b_1 b_4-b_2 b_3)\mathsf{D}
\big)(w)}{z-w}\,.
\end{align}
The order $1$ pole should be twice $T$ in \eqref{eqTansatzOd3}, so we can solve for the $a_i$ in terms of the $b_i$:
\begin{equation}
T=
(b_1^2+b_2^2)\mathsf{T}
+(b_3^2 + b_4^2)\no{\mathsf{J^3J^3}}
+(b_1 b_3 + b_2 b_4)\mathsf{C}
+(b_1 b_4-b_2 b_3)\mathsf{D} \,.
\end{equation}
The order 3 pole in the last OPE fixes the central charge, $c=9 (b_1^2 + b_2^2) - 6 (b_3^2 + b_4^2)$.

Let us calculate the $TG$ OPE and compare the order $2$ pole with $3G/2$ in \eqref{eq:GansatzOd3}. We obtain the equation
\begin{equation}
\left(-3 a_1+6 a_2+1\right) \left(b_3\mathsf{A} +b_4 \mathsf{B}\right)=\left(a_1-2 a_2-1\right) \left(b_1 \mathsf{G}+b_2 \mathsf{G^3}\right)\,.
\end{equation}
Recall that $a_1=b_1^2+b_2^2$ and $a_2=b_3^2+b_4^2$. There are two interesting possibilities:
\begin{itemize}
\item \textbf{Option 1: $a_1-2 a_2-1= 0$}
and then $-3 a_1+6 a_2+1 \neq 0$, which implies $b_3\mathsf{A} +b_4 \mathsf{B}=0$, which is only possible if $b_3=b_4=0$. We then get $b_3^2+b_4^2=a_2=0$, and thus $a_1=1$. Without loss of generality, $b_1=\cos\theta$, $b_2=\sin\theta$ and we end up with
\begin{align}
T&=\mathsf{T}\,, \qquad\qquad c=9\,,\\
G&=\cos\theta\,\mathsf{G}+\sin\theta\,\mathsf{G^3}\,.
\end{align}
This is the $\mathcal{N}=1$ superconformal algebra expected by proposition 1.1 (p.\,\pageref{prop:1}).
\item \textbf{Option 2: $-3 a_1+6 a_2+1= 0$} and then $a_1-2 a_2-1\neq 0$, which implies $b_1 \mathsf{G}+b_2 \mathsf{G^3}=0$, which is only possible if $b_1=b_2=0$. We then get $b_1^2+b_2^2=a_1=0$, and thus $a_2=-1/6$. Without loss of generality, $b_3={{i}}\cos\theta/\sqrt{6}$, $b_4={{i}}\sin\theta/\sqrt{6}$ and we end up with
\begin{align}
T&=\mathsf{T}^\triangleleft=-\frac{1}{6}\no{\mathsf{J^3J^3}}\,, \qquad\qquad c=1\,,\\
G&=\mathsf{G}^\triangleleft= \frac{{{i}}}{\sqrt{6}}\big(\cos\theta\,\mathsf{A}+\sin\theta\,\mathsf{B}\big)\,.
\end{align}
This is enough to satisfy all the defining OPEs of the $\mathcal{N}=1$ superconformal algebra.
\end{itemize}
It is impossible to choose neither of these options. This follows from the fact that a general linear combination of $\mathsf{G}, \mathsf{G^3}, \mathsf{A}$, and $\mathsf{B}$ vanishes only if all coefficients vanish individually.

Choosing option 1 is to follow the route of proposition 1. Just like the $\mathcal{N}=2$ superconformal algebra, $\text{Od}^3$ contains a unique, up to scale weight $2$ superprimary with respect to the $\mathcal{N}=1$ algebra in option 1. Concretely, this can be proven by setting to zero the order $2$ pole in the OPE $GW$, where we take the ansatz \eqref{eq:WansatzOd3} for $W$. The result is $W$ exactly as in proposition 1. From here, the proof in the main text shows that $\mathsf{c}=9$ does not allow an embedding of FS.

Choosing option 2 does not improve the situation. Again taking the ansatz \eqref{eq:UansatzOd3} for $W$, the order $2$ pole in the OPE $GW$ leaves the unique possibility
\begin{equation}
W=c_1\left(\mathsf{T}+\frac{1}{6}\no{\mathsf{J^3J^3}}\right) \,.
\end{equation}
But this field is actually singular with respect to the $\mathcal{N}=2$ algebra generated by $\mathsf{J^3}^\triangleleft$, $\mathsf{G}^\triangleleft$, $\mathsf{G^3}{}^\triangleleft$ and $\mathsf{T}^\triangleleft$. In fact, the partner $U$ of $W$ (defined as the order $1$ pole in the OPE $GW$) is singular in the whole $\text{Od}^3$ algebra. It must be quotiented out for $\text{Od}^3$ to satisfy associativity. Also $\wick{\c {\mathsf{T^\triangleleft}}(z) \c W(w)}=0$, clearly contradicting the FS algebra.

\section{$\mathcal{SW}_{\hat{c}}(\frac{3}{2},2)$ mode algebra} \label{app:FS_ModeAlgebras}

The FS OPEs in section~\ref{sec:FSunitarity} translate to the graded commutators below:
\begin{align}
[L_m, L_n] &= (m-n)L_{m+n} + \frac{c}{12}m(m^2-1)\delta_{m+n,0}\,, \\
[L_n, G_i] &= \left(\frac{n}{2}-i\right)G_{n+i}\,, \\
\{G_i, G_j\} &= 2 L_{i+j} + \frac{c}{3} \left(i^2 - \frac{1}{4}\right) \delta_{i+j,0}\,,
\end{align}
\begin{align}
[L_m, W_n] &= (m-n)W_{m+n}\,,\\
[L_n, U_i] &= \left(\frac{3n}{2}-i\right)U_{n+i}\,,\\
[G_i, W_n] &= U_{i+n}\,,\\
\{G_i, U_j\} &= (3i-j)W_{i+j}\,,\\
\end{align}
\begin{align}
[W_m, W_n] &= (m-n)\left(\mu L_{m+n}-\nu W_{m+n}\right)+\frac{c\mu}{12}m(m^2-1)\delta_{m+n,0}\,,\\
[W_n, U_i] &= \big(\boxope{WU}_{\,1}\big)_{n+i} + (n+1) \Bigg(\left(\frac{3}{2}-\frac{n}{2}+i\right)\mu G_{n+i}-\nu U_{n+i}\Bigg) \,,\\
\{U_i,U_j\} &= \big(\boxope{UU}_{\,1}\big)_{i+j} - \left(i+\frac{3}{2}\right)\left(j+\frac{3}{2}\right)(2\nu W_{i+j}-5\mu L_{i+j}) - \frac{c\mu}{12}\left(i^2-\frac{9}{4}\right)\left(i^2-\frac{1}{4}\right) \delta_{i+j,0} \,.
\end{align}
\begin{align}
\big(\boxope{WU}_{\,1}\big)_{n+i} &= 54 (c-15) \no{TG}_{n+i}+54 \no{GW}_{n+i}\\
&\qquad\quad+(c-15)^2 \left(n+i+\frac{5}{2}\right)\left(n+i+\frac{3}{2}\right)G_{n+i}+2 (c+12) \left(n+i+\frac{5}{2}\right)U_{n+i} \\
\big(\boxope{UU}_{\,1}\big)_{i+j} &= 3 \Big[        (c-15)\big(  9 \no{G'G}_{i+j}+36 \no{TT}_{i+j}+(2c-3)(i+j+3)(i+j+2)L_{i+j}  \big)   \\
&\qquad\qquad\qquad   -18 \no{GU}_{i+j}+36 \no{TW}_{i+j}+(c-6) (i+j+3)(i+j+2)W_{i+j}\Big]
\end{align}

It remains to properly define modes of normal ordered products in terms of modes of generators. This is straightforward to do for the Neveu--Schwarz moding using
\begin{align}
\no{AB}_n = \sum_{m\leq -h_A} A_m B_{n-m}
+(-1)^{|A||B|}
\sum_{m>-h_A} B_{n-m} A_m \,. \label{eq:NOPmodes}
\end{align}
In the Ramond sector, there is an ambiguity with \eqref{eq:NOPmodes} when $A$ is fermionic since the ranges of the sums are not well-defined. This is addressed in \cite[appendix C]{Gepner:2001px}. The problematic cases for the FS algebra are
\begin{equation}
\no{GW}_i\,,\qquad\qquad
\no{G'G}_i\,,\qquad\qquad
\no{GU}_i\,.
\end{equation}
The problem is mild for the first one, since we can express $\no{GW}$ in terms of $\no{WG}$ and then use \eqref{eq:NOPmodes} because $W$ is bosonic.
\begin{align}
\no{GW}^{\text{R}}_i
&{\;}{{{=}}}{\;} \sum_{m\leq -2} W_m G_{i-m}
+
\sum_{m>-2} G_{i-m} W_m
-\left(i+\frac{5}{2}\right)U_i\,.
\end{align}
The other correct Ramond expressions are found to be \cite{Gepner:2001px}
\begin{align}
\no{GG'}^{\text{R}}_i
&{\;}{{{=}}}{\;}
\sum_{m\leq - 1} G_m (G')_{i-m}
-
\sum_{m > - 1} (G')_{i-m}G_m
+\frac{5c}{64}\delta_{i,0}
+\left(i+\frac{9}{4}\right)L_i \,, \\
\no{GU}^{\text{R}}_i
&{\;}{{{=}}}{\;}
\sum_{m\leq - 1} G_m U_{i-m}
-
\sum_{m > - 1} U_{i-m}G_m
+\frac{i+3}{2}
 W_i \,.
\end{align}

\newpage


\normalem

\bibliographystyle{alpha}

\bibliography{Bibliography_Fiset}

\end{document}